\documentclass[a4paper, amsfonts, amssymb, amsmath, reprint, showkeys, nofootinbib, twoside,floatfix]{revtex4-2}

\usepackage[english]{babel}
\usepackage[utf8]{inputenc}
\usepackage{array}
\usepackage{amsthm}
\usepackage{mathtools}
\usepackage{physics}
\usepackage{xcolor}
\usepackage{graphicx}
\usepackage[left=13mm,right=13mm,top=25mm,bottom=25mm,columnsep=15pt]{geometry} 
\usepackage{adjustbox}
\usepackage{placeins}
\usepackage[T1]{fontenc}
\usepackage{lipsum}
\usepackage{csquotes}

\usepackage{fancyhdr} 
\fancyheadoffset{0cm}

\fancypagestyle{plain}{%
  \fancyhf{}%
  \fancyhead[R]{\thepage}%
}

\usepackage{xr-hyper} 
\makeatletter
\usepackage[pdftex, pdftitle={Article}, pdfauthor={Author}]{hyperref}
\usepackage{epstopdf}
\usepackage{siunitx}
\usepackage{xcolor}
\usepackage{threeparttable}
\usepackage{tabularx}
\begin{document}
\raggedbottom

\title{Wideband integrated high-speed graphene-silicon slot-waveguide electro-absorption modulator at 2 \textmu m and 1.5 \textmu m wavebands}
\author{Chao Luan$^{1,2*}$}
\author{Deming Kong$^{1}$}
\author{Yunhong Ding$^{1*}$}
\author{Hao Hu$^{1*}$}

\affiliation{$^{1}$DTU Electro, Department of Photonics Engineering, Technical University of Denmark, DK-2800, Kgs. Lyngby, Denmark}

\affiliation{{$^{2}$Current address: Research Laboratory of Electronics, MIT, Cambridge, MA, 02139, USA}}

\affiliation{{$*$chaoluan@mit.edu; yudin@dtu.dk; huhao@dtu.dk}}

\begin{abstract}

The 2-\textmu m waveband, emerging as a highly promising candidate for optical communication, offers an extended wavelength window for high-speed optical transmission. Despite its potential, the development of integrated electro-optic (E/O) modulators operating at this wavelength range has been limited. Such E/O modulators are crucial for high-speed optical communication systems at the 2-\textmu m waveband. In this work, we propose and experimentally demonstrate high-performance E/O absorption modulators based on a graphene-silicon slot waveguide. Our approach enables wideband, high-speed, efficient, robust and compact modulators at both 2-\textmu m and 1.5-\textmu m wavebands. This work represents a significant advancement towards the realization of high-speed integrated E/O modulators for optical communication systems operating at the 2-\textmu m wavelength range. 

\end{abstract}

\maketitle
\section{Introduction}

The Information Age has witnessed a digital explosion, marked by an exponential growth of data at a rate of tenfold every four years, posing a looming capacity crunch in optical fiber networks in the next decade \cite{richardson2010filling, cisco2019vni,jasion2020hollow, Miller2017,Reed2010}. While solutions such as data compression or space division multiplexing techniques, such as multiple parallel optical links, can mitigate the issue to some extent, they are not indefinitely scalable due to increased power consumption \cite{desurvire2011science}. Therefore, there is a pressing need to explore novel technologies capable of providing scalable and energy-efficient solutions for large-capacity optical communication, effectively addressing the challenges posed by the growing data demands of the digital era.

A promising solution to alleviate the capacity crunch in optical fiber networks is to explore the 2-\textmu m waveband, due to the availability of low-loss, low-nonlinearity, hollow-core photonic bandgap fiber and ultra-broadband thulium-doped fiber amplifier (TDFA) \cite{cao2018highspeed2um, kong2022superbroadband, desurvire2011science, poletti2013towards, slavik2015ultralow, luan2022integrated, sakr2020interband, roberts2005ultimate, petrovich2025broadband, li2013diode, zhao2025ultra, kuznetsov2025ultra}. Moreover, the 2-\textmu m waveband is compatible with silicon photonics, enabling fully integrated high-performance transceivers that can be mass-produced using mature CMOS processes. Various silicon photonic devices for the 2-\textmu m waveband have been demonstrated, including low-loss stripe and rib waveguides, grating couplers, mode converters, arrayed waveguide gratings, and photodetectors \cite{liu2020thermooptic, zheng2019fourmode, sia2020selector, shen2021dualmode, guo2020graphenepd,yin2019bp2um, hagan2020tdfa}. Despite these advancements, there have been limited demonstrations of high-speed integrated 2-\textmu m waveband E/O modulators \cite{shen2022bistability, pan2021lnoi2um, li2026ultra, hagan2020tdfa, wang2021high, wang2024efficient}. So far, silicon E/O modulators at 2-\textmu m waveband rely on changing carrier concentration in silicon, which is an intrinsically weak effect and usually have limited bandwidth \cite{shen2022bistability,hagan2020tdfa, wang2021high}. Another approach involves thin film LiNbO$_3$ on insulator (TFLNOI) modulator based on the Pockels effect \cite{pan2021lnoi2um, li2026ultra}. However, the fabrication process of TFLNOI modulator is not CMOS compatible. In addition, both silicon and TFLNOI integrated modulators are based on the Mach–Zehnder interferometer (MZI) or micro-ring structure, which results in either large footprints or limited working wavelength range and high temperature sensitivity. A recently demonstrated TFLNOI Mach-Zehnder modulator has achieved broad spectral coverage spanning the O-band through the 2~\textmu m waveband with a 50 GHz bandwidth, but needs a large footprint of 9~mm due to the weak phase modulation efficiency \cite{li2026ultra}. Instead, E/O absorption modulators achieving efficient and non-resonant flat operation, which is not limited by the wavelength and fabrication error sensitivity of microring structures or the extensive device length of travelling-wave millimetre-scale MZIs. Given the importance of E/O absorption modulators at the 2-\textmu m waveband for optical transmission systems and interconnects in data-centers, there is a need to explore efficient, large-bandwidth, robust and compact solutions for 2-\textmu m waveband E/O absorption modulators. However, the pool of materials exhibiting electrically tuneable optical absorption is remarkably limited in nature. Conventional electro absorption mechanisms such as the Franz-Keldysh effect and the quantum-confined Stark effect operate only near the material's bandgap edge, restricting each semiconductor to a narrow spectral window and leaving few viable candidates at 2~\textmu m waveband.


\begin{figure*}[t!]
\centering
\includegraphics[width=0.68\textwidth]{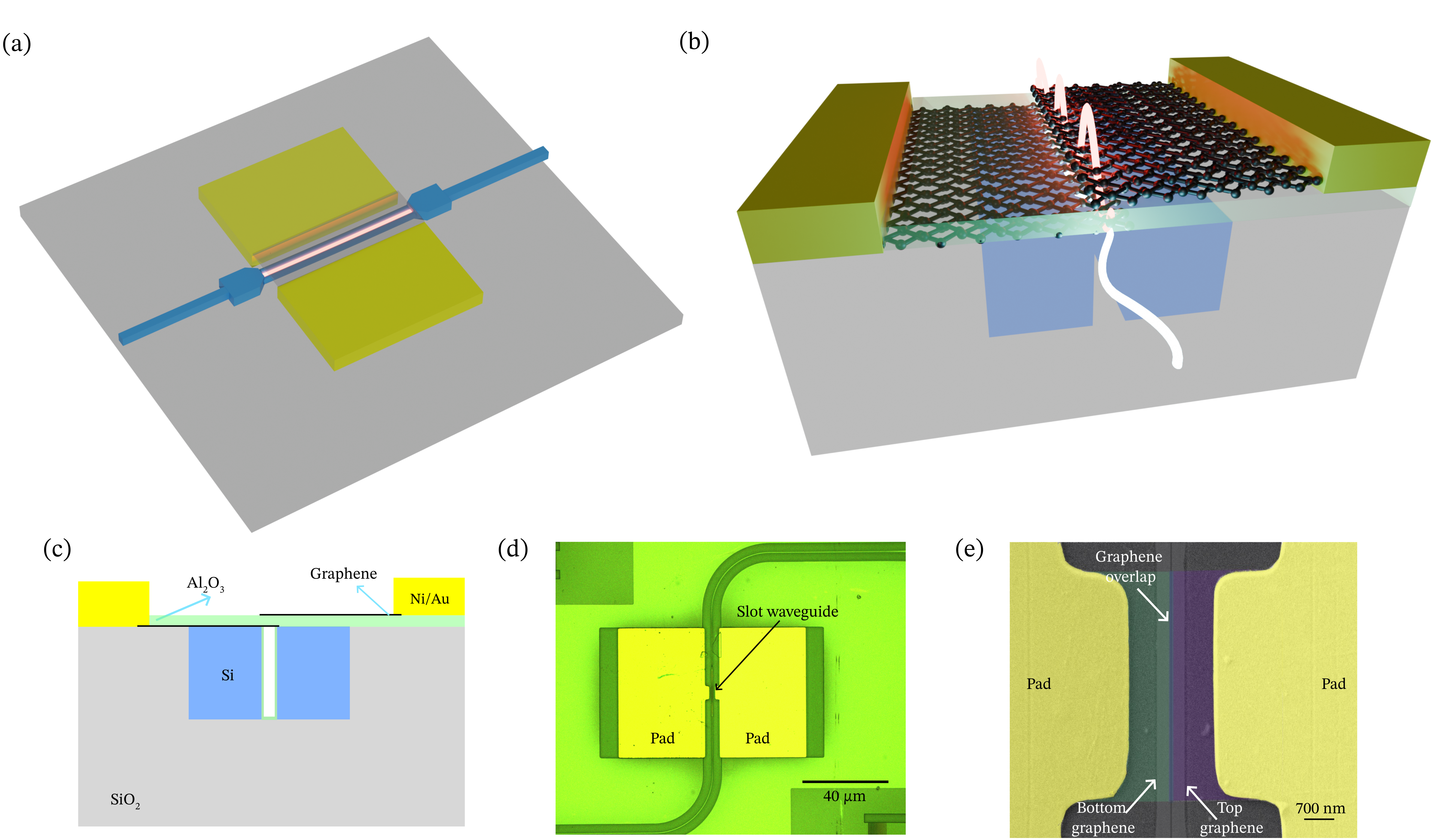}
\caption{Device concept. (a-c) Schematic and cross section of the graphene-silicon slot-waveguide E/O absorption modulator. (d) Optical microscope image of the graphene-silicon slot waveguide E/O absorption modulator. (e) SEM image of the graphene-silicon slot waveguide E/O absorption modulator. }
\label{fig:1}
\end{figure*}

Graphene is a promising optoelectronic material for E/O modulators due to its unique properties \cite{liu2011graphenebroadband, liu2012doublelayer, giambra2019doublelayer, dalir2016athermal, hu2016broadband10g, agarwal2021hbn, mohsin2014lowloss, wang2009broadband,sorianello2018phase, ono2020ultrafast, luan20202,luan2022high, luan2022highpdp, luan2022integrated, luan2022ultra, luan20202}. Featuring a gapless band structure, graphene exhibits electrically tuneable broadband optical absorption spanning from UV to THz wavebands. Its fast photocarrier generation/relaxation time, in the scale of femtoseconds, enables the realization of large bandwidths reaching up to 500 GHz. Graphene also exhibits low intrinsic optical loss, strong light-matter interaction, and CMOS compatibility, making it attractive for integrated optoelectronic devices. Several graphene E/O modulators at the 1.5-\textmu m waveband have been demonstrated \cite{liu2011graphenebroadband, liu2012doublelayer, giambra2019doublelayer, dalir2016athermal, hu2016broadband10g, agarwal2021hbn, mohsin2014lowloss, wang2009broadband,sorianello2018phase, ono2020ultrafast, luan20202,luan2022high, luan2022highpdp, luan2022integrated, luan2022ultra, luan20202}. Among these, the graphene-dielectric-graphene configuration is widely adopted. This configuration leverages the high mobility and low loss of graphene while avoiding the complicated ion implantation process in silicon. However, the atomic-thin nature of graphene poses challenges, leading to limitations in light-graphene interaction and modulation efficiency, or necessitating a large device footprint for high-speed modulators, which is often undesirable. Notably, despite graphene's broad working wavelength range and its demonstrated ability to modulate light at the 2-\textmu m waveband, graphene E/O modulators at this wavelength remain unrealized. This gap underscores the necessity for further research and development efforts to unlock the potential of graphene for high-performance E/O modulators at the 2-\textmu m waveband.

In this paper, we propose and demonstrate a broadband, high-speed, efficient, robust, and compact graphene-silicon electro-absorption modulator, utilizing a deep-subwavelength silicon slot waveguide with a compact footprint of only 6-\textmu m. Such a compact footprint only needs the lumped electrode, which avoids the waveguide dispersion and carrier velocity mismatch for the travelling-wave electrode when multi-wavebands data are modulated. This makes the modulator works at broad wavebands covering both 2-\textmu m and 1.5-\textmu m with the same active region. When operating at the 2-\textmu m waveband, our silicon-graphene slot waveguide modulator exhibits a measured 3-dB bandwidth beyond 40 GHz (limited by the measurement setup), a modulation efficiency of 0.22 dB/\textmu m, and a low insertion loss of 0.6 dB. Moreover, our graphene modulator also has a superior performance at the 1.5-\textmu m waveband, exhibiting a 3-dB bandwidth exceeding 70 GHz (limited by the measurement setup) and a modulation efficiency of 0.2 dB/\textmu m.

\section{Device structure}

\begin{figure*}[t!]
\centering
\includegraphics[width=0.8\textwidth]{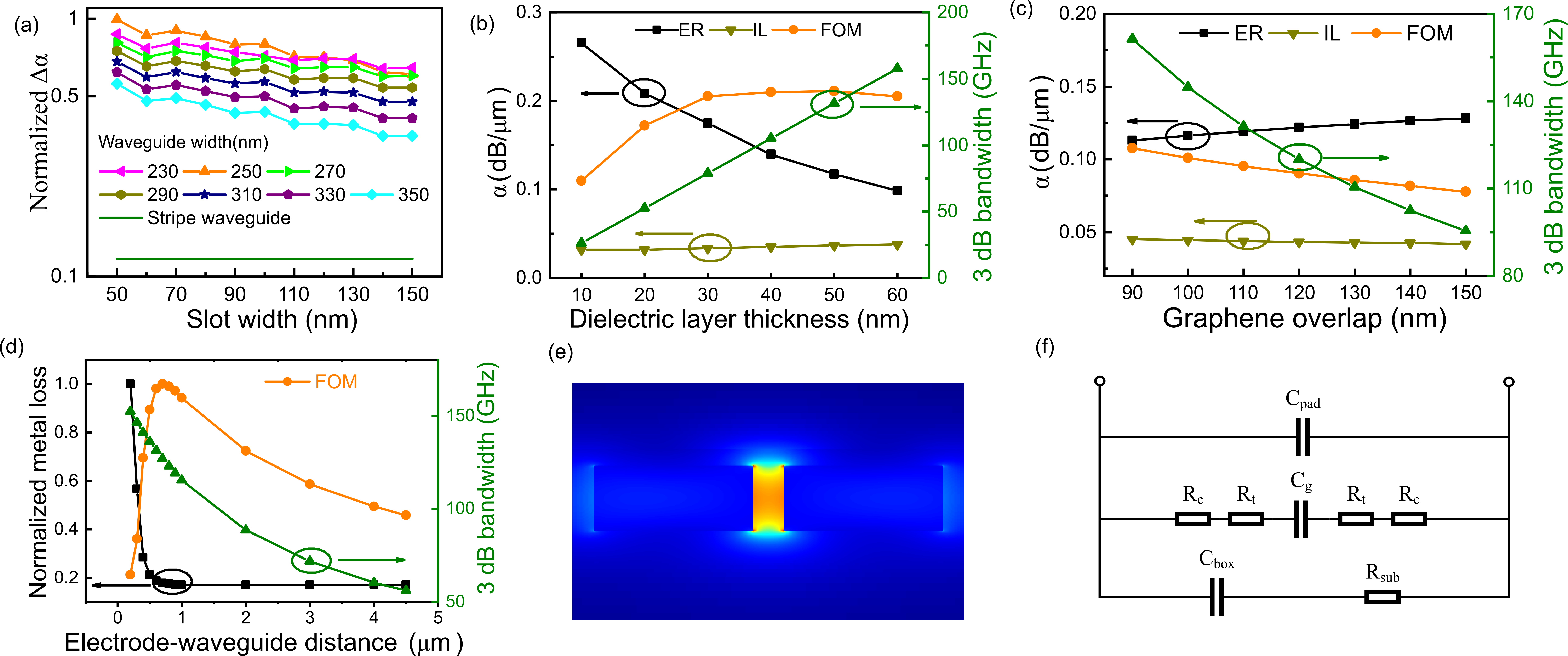}
\caption{Double-layer-graphene slot-waveguide E/O modulator optimization. (a) Calculated and normalized absorption coefficient change $\Delta a$ versus the slot waveguide dimension. (b) Calculated normalized extinction ratio, bandwidth, insertion loss and modulation efficiency-bandwidth product of the slot waveguide modulator versus dielectric layer thickness. The highest modulation efficiency-bandwidth product was obtained at the dielectric layer thickness of 50 nm. (c) Calculated normalized extinction ratio, bandwidth, insertion loss and modulation efficiency-bandwidth product of the slot waveguide modulator versus double layer graphene overlap width. (d) Calculated normalized metgal absorption loss, modulation bandwidth and modulation efficiency-bandwidth product of the slot waveguide modulator versus electrode to waveguide distance. (e) Calculated eigenmode of the slot waveguide structure. (f) RC circuit model of the slot waveguide modulator.}
\label{fig:2}
\end{figure*}

Fig.~\ref{fig:1} (a), (b) and (c) depict the schematic of the proposed double-layer graphene slot-waveguide modulator. The slot waveguide consists of two silicon strip waveguides, each with a width of 250~nm, and separated by a 50-nm-wide air slot to achieve the largest absorption coefficient change $\Delta \alpha$. As indicated, the slot waveguide exhibits a much larger $\Delta \alpha$ compared to the stripe waveguide, attributed to the stronger light-graphene interaction in the slot waveguide. Fig.~\ref{fig:2} (e) shows the calculated eigenmode of the slot waveguide, indicating that the electric field is tightly confined ($>45\%$) in the air slot. This extreme light confinement around graphene significantly enhances the light--graphene interaction and modulation efficiency. This facilitates a high modulation efficiency even with a short graphene length of only 6-\textmu m. Additionally, Fig.~\ref{fig:2} (f) shows the equivalent resistance-capacitance (RC) circuit of the double-layer graphene slot-waveguide electro-absorption modulator. The double-layer graphene slot-waveguide modulator has a graphene overlap of only 110~nm and a dielectric layer thickness of 50~nm. The small partial overlap of graphene and the thick dielectric layer reduce the capacitance and enable a large modulation bandwidth. Moreover, because this device operates as an electro-absorption modulator rather than a resonant modulator, its operating wavelength range is not fundamentally constrained by a cavity resonance. This is advantageous for broadband operation and also makes the device more robust against temperature variations and fabrication error. Therefore, employing a slot waveguide with a small graphene overlap enables large bandwidth, high modulation efficiency, low energy consumption, improved robustness, and broadband operation within a compact device footprint. 

\section{Device design}

The slot waveguide modulator mainly comprises three distinct structural sections. The first is the silicon strip waveguide, which serves as the light propagation region. The second is the deep sub-wavelength silicon slot waveguide region where light modulation happens, and the third is the strip-to-slot waveguide mode converter where a 1×2 multimode interferometer (MMI) is employed. The MMI dimensions is 1.48 \textmu m in width and 1.58 \textmu m in length, which has low loss at 2 \textmu m waveband. Based on the self-imaging principle, the input field is reproduced as two images at the output of the MMI. A second, reversed MMI is placed at the opposite end of the slot section to convert the slot mode back to the strip mode. Fully etched photonic crystal grating couplers at 1.5 \textmu m and 2 \textmu m wavelength bands are designed to couple light into and out of the chip. The grating couplers are designed to have a coupling angle of 8$^{\circ}$ to match the commercial fiber array. Non-uniformed photonic crystal hole radius and pitch period are designed to maximize the coupling efficiency and achieve apodized coupling.

The optimization of the slot width and waveguide width is crucial for achieving a high extinction ratio, as illustrated in Fig. 2(a). The extinction ratio is enhanced with decreasing slot width, due to the field concentration effect. However, considering the fabrication yield, a slot width of 50 nm was selected for the experimental demonstration. In the simulation, the highest extinction ratio is achieved with a waveguide width of 250 nm, exhibiting over an 8-fold enhancement compared to the stripe waveguide. This emphasizes the significant improvement in terms of the extinction ratio achieved by using the slot waveguide.

In addition to the extinction ratio, modulation bandwidth is a crucial parameter for a high-performance modulator. Previous approaches relied on a thick gate oxide to increase bandwidth, resulting in a substantial overlap width of the double-layer graphene and a large graphene footprint, consequently leading to high device capacitance (in the order of  tens of femtofarads), or introduce high-$\kappa$ dielectrics such as hBN or HfO$_2$, which offer a thinner gate barrier while maintaining low leakage current \cite{agarwal2021hbn, tiberi2025graphene}. In contrast, our silicon slot waveguide design enables a significant improvement in bandwidth through partial graphene overlap and a compact device footprint. Key parameters such as the thickness of the dielectric layer, the overlap width of the double-layer graphene, and the electrode-to-waveguide distance play pivotal roles in modulator bandwidth. Through careful optimization of these parameters within the slot waveguide configuration, we achieve a substantial increase in the modulation bandwidth, presenting a promising solution for high-speed electro-optic modulation.

\begin{figure*}[t!]
\centering
\includegraphics[width=0.8\textwidth]{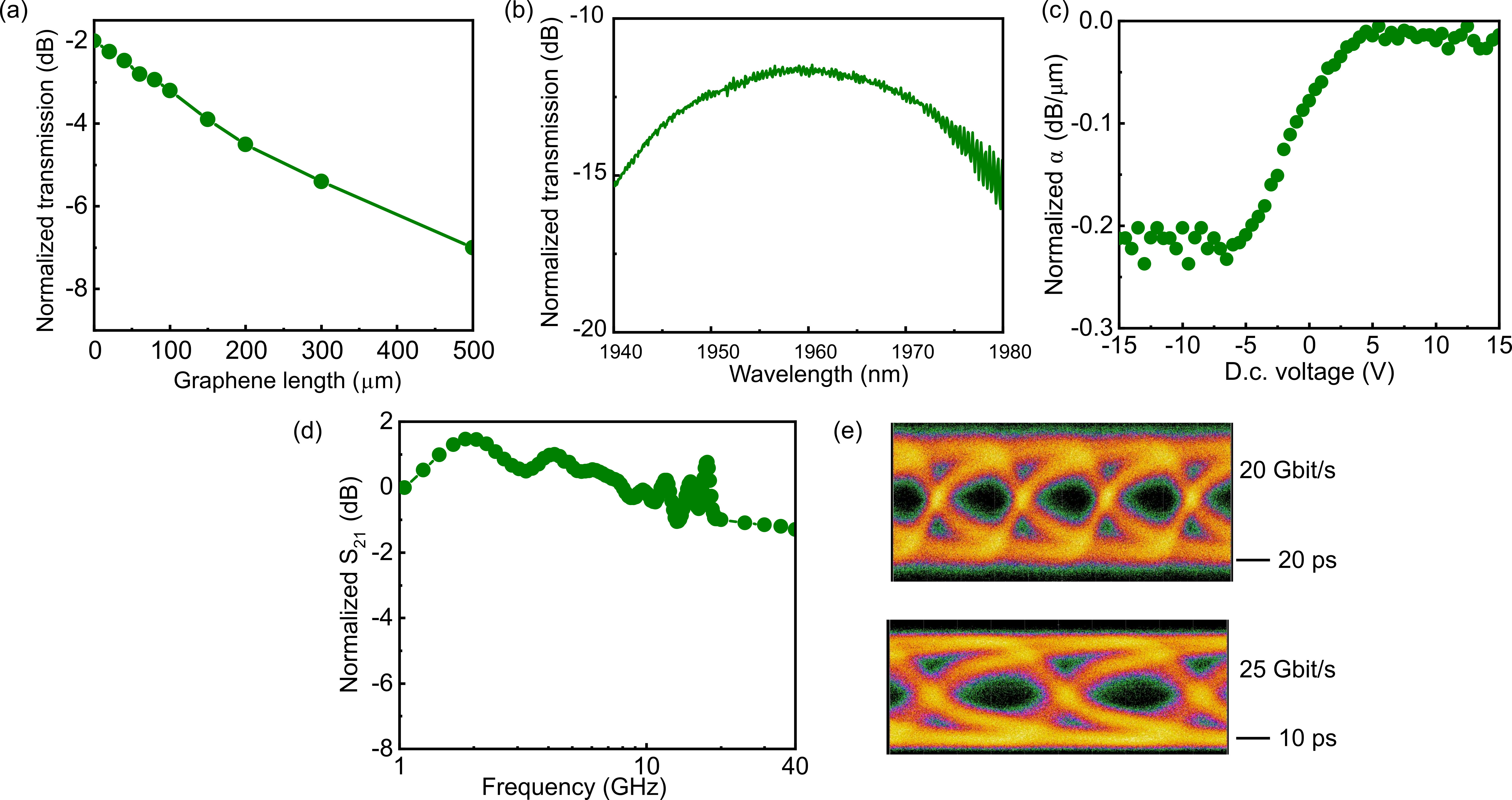}
\caption{Characterization of the double-layer-graphene slot-waveguide modulator at 2\textmu m wavelength bands. (a) Optical transmission of the 2\textmu m waveband graphene modulator with different graphene length. (b) Optical transmission of the 2\textmu m waveband grating couplers. (c, d) Measured absorption coefficient of the graphene silicon slot waveguide modulator at 2\textmu m waveband. (e) Measured electro-optic S$_{21}$ frequency response of the modulator at 2\textmu m wavelength bands. The bandwidth of the modulator is beyond 22 GHz. (e). Measured eye diagram of the modulator at 2\textmu m wavelength bands at 25 Gbit/s and 20 Gbit/s.}
\label{fig:3}
\end{figure*}

Fig.~\ref{fig:2}(b) illustrates the calculated extinction ratio (ER), insertion loss (IL), modulation bandwidth (BW), and figure of merit (FOM = ER $\cdot$ BW / (V$_{\mathrm{pp}} \cdot$ IL $\cdot$ L)) of the graphene-silicon electro-absorption modulator as a function of the dielectric layer thickness. As the dielectric layer thickness increases from 10 nm to 60 nm, the modulation bandwidth increases from 30 GHz to 160 GHz, while the extinction ratio reduces from 0.26 dB/\textmu m to 0.1 dB/\textmu m, all at a driving voltage of 6 V. The highest figure of merit is achieved at a dielectric layer thickness of 50 nm.
Fig.~\ref{fig:2} (c) illustrates the extinction ratio, bandwidth, insertion loss, and figure of merit of the graphene E/O modulator as the graphene overlap width varies from 90 nm to 150 nm. The modulation bandwidth decreases from 160 GHz to 95 GHz, while the extinction ratio only increases slightly from 0.12 dB/\textmu m to 0.13 dB/\textmu m due to the field concentration effect around the slot region. Therefore, a narrower overlap is favored to achieve a higher figure of merit. Considering the constraints of fabrication yield, we selected the double-layer graphene overlap width of 110 nm. 
The distance between the electrode and waveguide is also optimized to maximize modulation bandwidth while avoiding excessive loss. As shown in Fig.~\ref{fig:2}(d), the modulation bandwidth increases as the distance between the electrode and waveguide decreases. However, there is undesirable large metal absorption when the electrode-to-waveguide distance is smaller than the evanescent decay length. Fortunately, due to the tight light confinement in the slot waveguide, a small electrode-to-waveguide distance of 700 nm can be achieved without introducing high loss. These findings highlight the effectiveness of the proposed design in achieving a balance between key parameters for the electro-optic modulator.

In addition to the optical waveguide geometry, the electrical design of the modulator was systematically optimized. The electrode follows a lumped-element approach, and the electrical design was optimized through the following strategies: (i) thick oxide underneath the pads—the silicon beneath the pads is etched away, and a high-resistivity SOI platform is used to reduce the pad-to-substrate parasitic capacitance; (ii) minimized routing metal length—the pads are connected directly to the graphene contact points, eliminating additional parasitic capacitance from long interconnect traces; (iii) small graphene–metal contact overlap of 700 nm to minimize parasitic capacitance at the contact interface; (iv) close electrode-to-waveguide placement, enabled by the strong optical confinement within the slot, which reduces the graphene sheet resistance in the access path and thereby lowers the total series resistance; and (v) rapid thermal annealing (RTA) to reduce the graphene–metal contact resistance. These design choices collectively minimize the RC time constant and enable the large electro-optic bandwidth of the modulator.

\section{Device fabrication and characterization}

The graphene-silicon electro-optic modulator was fabricated on a high-resistivity 220-nm silicon-on-insulator (SOI) platform. The fabrication process involved several steps, including e-beam lithography, induced chemical plasma (ICP) etching, atomic layer deposition (ALD), e-beam evaporation, lift-off, and graphene wet transfer. Further details on the fabrication process are provided in the Methods section. The contact resistivity was reduced by using RTA, which led to a reduction of approximately 68\% in the contact resistance. A newly developed wet transfer process was employed to transfer a $2\,\text{cm} \times 2\,\text{cm}$ single-layer graphene sheet onto the chip.

Fig.~\ref{fig:1} (d) presents an optical microscopy image of the fabricated device including the slot waveguide and the electrodes. Fig.~\ref{fig:1} (e) shows an artificial-colored scanning electron microscopy (SEM) image of the slot waveguide. The top (purple) and bottom (green) graphene layers form a capacitor through partial overlap, which is designed to reduce device capacitance and enhance light-graphene interaction. The 1.5 \textmu m waveband and 2 \textmu m waveband modulators have the same optical microscope and SEM images except for the grating couplers.

The modulator (Device A) was first characterized at 2 \textmu m waveband. The apodized grating couplers, 14 \textmu m in width with a 400 \textmu m long adiabatic taper, exhibit an insertion loss of 6 dB at the 2 \textmu m waveband. The extra loss in the grating couplers is attributed to residual metal contamination during the graphene wet transfer process and Al$_2$O$_3$ residual falling into the photonic crystal holes. Additionally, the mode converter between the stripe waveguide and the slot waveguide has a measured loss of 0.3 dB at the 2-\textmu m waveband. Fig.~\ref{fig:3} (c) shows the transmission spectrum of the graphene modulator with DC bias voltages ranging from -15 V to 15 V, revealing a static modulation efficiency of 0.22 dB/\textmu m. The bias voltage was not increased further to avoid potential damage to the device.

\begin{figure*}[t!]
\centering
\includegraphics[width=0.8\textwidth]{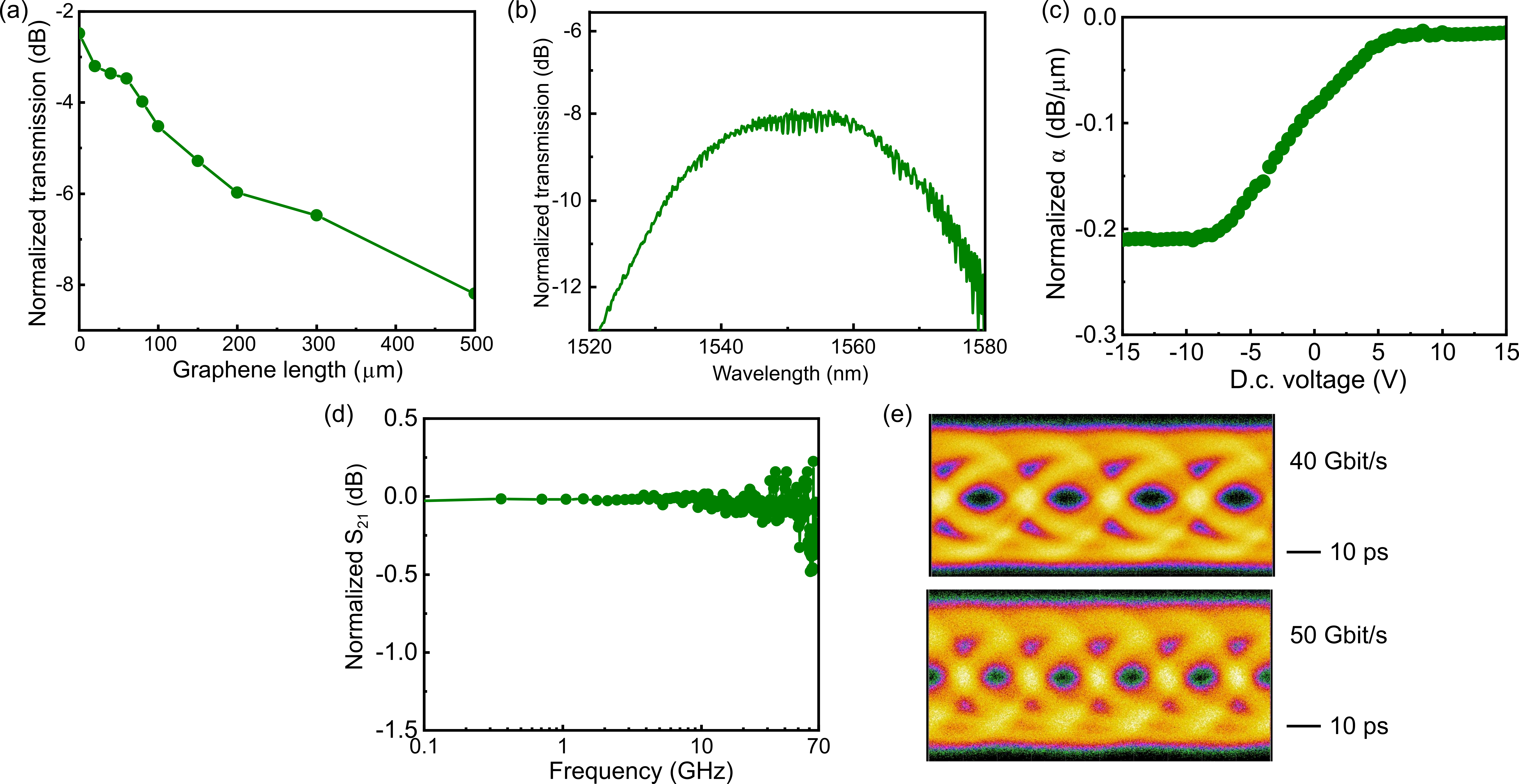}
\caption{(a) Optical transmission of the 1.5\textmu m waveband grating couplers. (b) Measured absorption coefficient of the graphene silicon slot waveguide modulator at 1.5\textmu m waveband. (c) Measured electro-optic S$_{21}$ frequency response of the modulator at 1.5\textmu m wavelength bands. The bandwidth of the modulator is beyond 70 GHz. (e). Measured eye diagram of the modulator at 1.5\textmu m wavelength bands at 40 Gbit/s and 50 Gbit/s.}
\label{fig:4}
\end{figure*}

Traditionally, achieving high modulation efficiency involves making trade-offs with bandwidth. However, the silicon slot waveguide design offers the unique advantage of simultaneously providing a large bandwidth and high modulation efficiency. The modulation bandwidth of the device was measured by a vector network analyzer (VNA) and a 2-\textmu m waveband photodetector. A continuous wave (CW) light generated by a distributed-feedback (DFB) laser at 2000 nm was launched into the modulator, and the modulated signal was received by the 2-\textmu m waveband photodetector (discovery DSC2-50S, 15-GHz bandwidth) and sent back to the VNA for bandwidth measurement. In Fig.~\ref{fig:3} (d), the measured frequency response of the modulator reveals a modulation bandwidth exceeding 40 GHz. It is noted that the measurement is limited by the bandwidth data of the equipment used in the experiment, electro optic bandwidth before 20 GHz was measured using photodetector, from 20~GHz to 40 GHz, we adopted an optical spectrum analyzer sideband method.

To further validate the large bandwidth performance of the graphene E/O modulator, we characterized another modulator (Device B) on the same chip. Device B has the same slot width, waveguide width and electrical structure as device A, but it includes a grating coupler designed for the 1.55-\textmu m waveband, which allows for high-speed measurements (Finisar XPDV3120R, 70-GHz bandwidth, Keysight N5227B VNA, 70-GHz bandwidth) in the lab. The measured losses of the grating coupler and mode converter at the 1.55-\textmu m waveband were 4 dB and 0.45 dB, respectively. Device B demonstrated a modulation efficiencies of 0.2 dB/\textmu m. In comparison, the state-of-the-art graphene electro-absorption modulator at the 1.55-\textmu m waveband achieved comparable modulation efficiency using 2D-3D integration of different dielectric materials, with bandwidth of 40 GHz. In contrast, our graphene slot-waveguide modulator at the 1.55-\textmu m waveband demonstrated a measured bandwidth beyond 70 GHz, as shown in Fig.~\ref{fig:4} (d), with the measurement limited by the bandwidth of the VNA used in the experiment. Because the 2-\textmu m waveband modulator has the same structure as the 1.55-\textmu m waveband modulator, its bandwidth is also expected to exceed 70 GHz. According to numerical calculations, the bandwidth of the graphene slot-waveguide modulator can potentially reach up to 150 GHz, thanks to the very small capacitance of the device.

The high-speed performance of our graphene slot-waveguide modulator was also supported by non-return-to-zero (NRZ) eye diagram measurements. High-speed data was generated using a bit pattern generator with a 2$^{15}$-1 pseudo-random binary sequence (PRBS) at 50~Gbit/s, which was then amplified to 2-V. At 2 \textmu m waveband, open eye diagram of 25\,Gbit/s was obtained, which is limited by the photodetector bandwidth, the resulting open eye diagram was shown in Fig.~\ref{fig:3} (e). In this configuration, our modulator exhibited a low power consumption of approximately 10~fJ/bit (CV$^2$/4). While at 1.55 \textmu m waveband, open eye diagram at 50 Gbit/s was demonstrated (setup limited). These results collectively indicate the high performance of our graphene slot-waveguide E/O modulator, highlighting its potential for high-speed optical communications.

\begin{figure}[t!]
\centering
\includegraphics[width=0.35\textwidth]{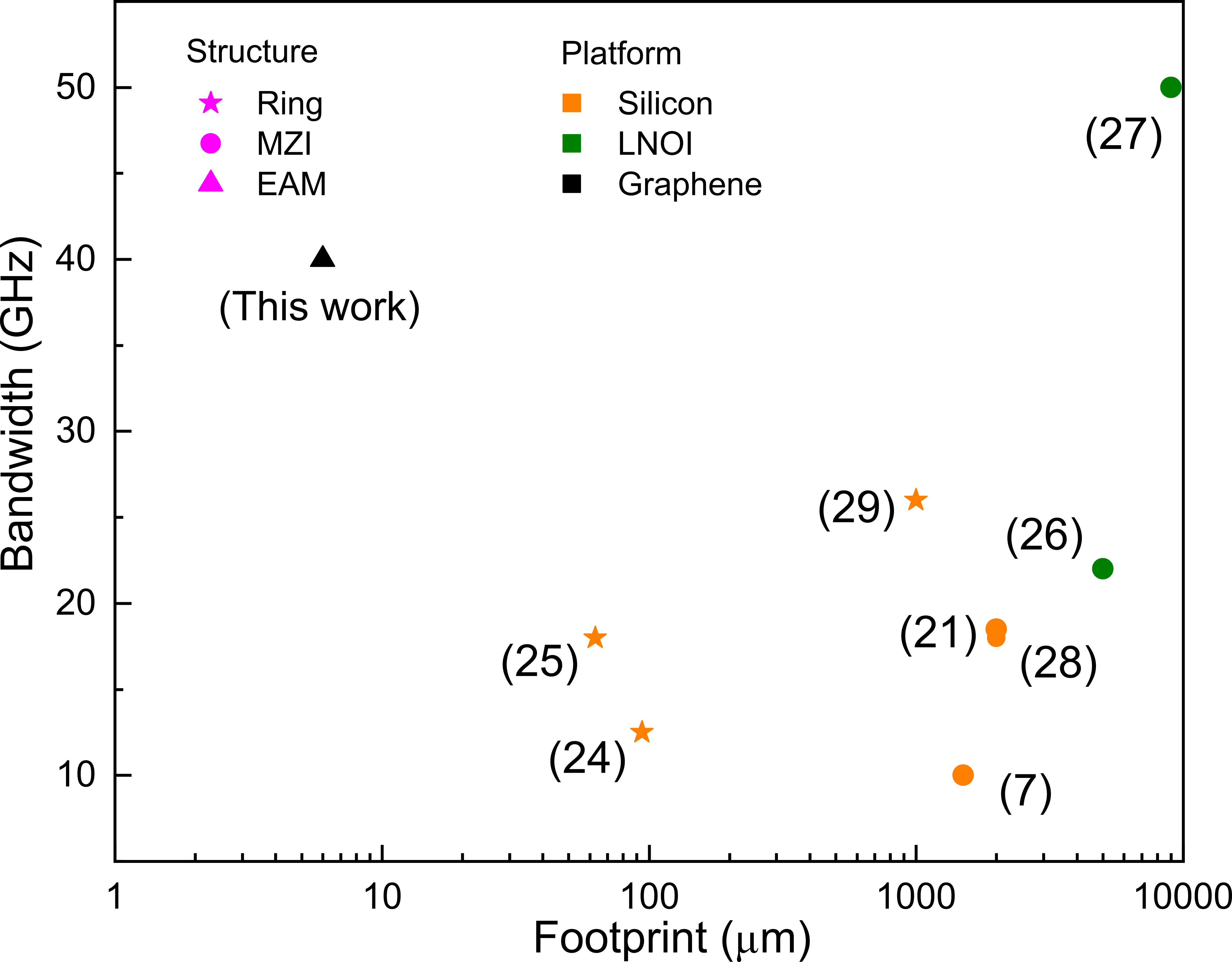}
\caption{Comparison of state-of-the-art E/O modulators at 2-\textmu m waveband. }
\label{fig:5}
\end{figure}

In comparison with state-of-the-art E/O modulators at the 2 \textmu m waveband, our graphene slot-waveguide modulator exhibits superior performance across various key parameters, including modulation bandwidth, footprint, and modulation efficiency. In addition, this is the first demonstration of electro absorption modulators at 2 \textmu m waveband. For instance, a silicon MZI modulator, as detailed in \cite{cao2018highspeed2um}, achieved an extinction ratio of 5.8 dB with a data rate of 20 Gbit/s. Similarly, a LiNbO$_3$ MZI modulator reported a modulation bandwidth exceeding 50 GHz. However, these modulators necessitate long device lengths, leading to larger footprints, high energy consumption, and integration challenges with other photonic components. Another reported micro-ring modulator at the 2 \textmu m waveband demonstrated a modulation bandwidth of 18 GHz but required precise wavelength alignment to the laser and exhibited vulnerability to temperature variation and fabrication error, mandating additional heaters for wavelength tuning and increasing energy consumption. In contrast, our graphene slot-waveguide modulator does not require wavelength alignment to the laser, is thus robust to temperature variation, and features a compact device length of only 6-\textmu m with a setup-limited modulation bandwidth exceeding 40 GHz (Fig.~\ref{fig:5}). This results in a high modulation efficiency of 0.22 dB/\textmu m with the calculated bandwidth over 150 GHz, indicating superior performance compared to other reported modulators at the 2 \textmu m waveband.

\section{Conclusion}
We have demonstrated integrated graphene-silicon slot-waveguide electro absorption modulators at both 2-\textmu m and 1.55-\textmu m wavebands, featuring large bandwidth, low loss, compact footprint, robust operation, and high modulation efficiency. These advantages are enabled by combining the slot-waveguide structure and the partial overlap of double-layer graphene, which simultaneously enhances the interaction between light and graphene and increases the modulation bandwidth by reducing the capacitance.

In our experiments, the fabricated modulator exhibited a measured modulation bandwidth exceeding 40 GHz at the 2 \textmu m waveband and exceeding 70 GHz at the 1.55 \textmu m waveband, demonstrating high-speed performance at both wavebands. It’s noted that these measurements are limited by our experimental setup. However, our calculation show that the bandwidth of the graphene slot-waveguide modulator can potentially extend up to 150 GHz, due to the very small capacitance of the device. These results highlight the potential of the integrated graphene E/O modulator for future high-speed optical communication systems, serving as a key component bridging electronics and photonics. Furthermore, our graphene slot-waveguide modulator demonstrated consistent and high modulation efficiency over a broad wavelength range. The almost constant modulation efficiencies at different wavelength bands highlight the versatility and potential of graphene-based modulators for diverse applications in photonic integrated circuits (PICs) and optical communication networks \cite{luan2026single, Miller2017, luan2025demonstration}. 

In summary, the integrated graphene-silicon slot-waveguide E/O modulators demonstrated in our work exhibit compelling performance characteristics, paving the way for practical applications of graphene in high-speed 2-\textmu m waveband optical communication systems and other advanced photonic devices.

\section*{Methods}

The graphene modulator was fabricated on a commercial high resistivity silicon-on-insulator (SOI) chip with a 220-nm-thick silicon layer on top of a 2-\textmu m-thick SiO$_2$ buried layer. E-beam lithography and inductively coupled plasma etching are used to fabricate the 50-nm-wide slot waveguides and other passive components. To ensure a high graphene transfer-quality yield, PECVD SiO$_2$ was deposited and planarized to the top surface of the waveguide by using standard chemical mechanical polishing (CMP) technique to provide a flat surface which otherwise tends to break the graphene across the waveguide edge while drying. A spacer of 5\textasciitilde7-nm-thick Al$_2$O$_3$ was then uniformly atomic layer deposited on the surface of the waveguide to isolate the carrier transportation between the graphene and silicon waveguide. CVD graphene on copper foil was spin-coated with 1~\textmu m AZ5214E resist, baked at \SI{90}{\degreeCelsius} for 2~minutes until dry, then soaked in a homemade copper etching solution (hydrochloric acid:DI water = 1:7, with a few drops of hydrogen peroxide) for 24~h and rinsed thoroughly in deionized (DI) water. To remove metal residuals, a RCA clean step-two solution was employed before the wet-transfer process. The transferred graphene was left to dry for one week. The graphene pattern was then defined using e-beam lithography and O$_2$ plasma etching. Next, the metal pads were defined using UV lithography and deposited by e-beam evaporation following a lift-off process with 10~nm Ni and 100~nm Au. A rapid thermal annealing (RTA) process was then used to significantly improve graphene--pad conductivity and reduce contact resistance. During the RTA process, the sample was ramped to \SI{450}{\degreeCelsius} in \SI{30}{\second} and held at \SI{450}{\degreeCelsius} for approximately \SI{1}{\minute} under a flowing gas mixture of 10\% hydrogen in nitrogen, with the cycle repeated five times. Direct deposition of high-dielectric-constant materials on pristine graphene by ALD is challenging due to the hydrophobic nature of the graphene basal plane. A 1~nm Al seed layer was thermally evaporated and quickly oxidized into Al$_2$O$_3$ upon exposure to air, after which 35~nm Al$_2$O$_3$ was deposited at \SI{200}{\degreeCelsius} by ALD. A similar process was performed for the top graphene layer and top metal pad as for the bottom layer. The top Al$_2$O$_3$ layer above the grating and metal pad was opened via H$_3$PO$_4$ wet etching.

\section*{Data Availability}
The data from this work is stored in the server computer (DTU-CZC8028T24) at DTU and will be made available upon reasonable request.

\renewcommand{\bibsection}{\section*{REFERENCES}}
\bibliography{main}

\section*{Acknowledgements}
The author thanks Dr. Yong Liu from DTU Electro for the helpful discussion during the deviec fabrication process, and Dr. Peixiong Shi, Dr Thomas Pedersen from DTU Nanolab for the equipment training. 

\subsection*{Author Contribution}

C.~Luan,  Y.~Ding and H.~Hu conceived the project. C.~Luan led the overall effort, including device design, simulation, fabrication, and characterization. C.~Luan also prepared the original manuscript. D.~Kong assisted with the high speed eye-diagram measurements at 1.5\textmu m waveband and 2\textmu m waveband. C.~Luan wrote the manuscript with the feedback from  all the authors. Y.~Ding and H.~Hu co-supervised the project. This work was supported by a research grant (15401) of Young Investigator Program (2MAC) and QUANPIC (00025298) from VILLUM FONDEN.

\section*{Competing Interests}
The author declares no competing interests.

\end{document}


\title{Supplementary Information: Wideband integrated high-speed graphene-silicon slot-waveguide electro-absorption modulator at 2 \textmu m and 1.5 \textmu m wavebands}

\author{Chao Luan$^{1,2*}$}
\author{Deming Kong$^{1}$}
\author{Yunhong Ding$^{1*}$}
\author{Hao Hu$^{1*}$}
\affiliation{$^{1}$DTU Electro, Department of Photonics Engineering, Technical University of Denmark, DK-2800, Kgs. Lyngby, Denmark}

\affiliation{\textcolor{black}{$^{2}$Current address: Research Laboratory of Electronics, MIT, Cambridge, MA, 02139, USA}}

\affiliation{\textcolor{black}{chaoluan@mit.edu; yudin@dtu.dk; huhao@dtu.dk}}

\maketitle
\tableofcontents
\newpage

\section{Modeling Graphene}

\subsection{Graphene surface conductivity material model}

The surface conductivity model is used in our numerical calculation. The surface conductivity of graphene can be expressed as:
\begin{align}
\sigma(\omega,\mu_c,\tau)
&=\sigma_{\mathrm{intra}}(\omega,\mu_c,\tau)+\sigma_{\mathrm{inter}}(\omega,\mu_c,\tau)
\tag{1.1}\\[4pt]
\sigma_{\mathrm{intra}}(\omega,\mu_c,\tau)
&=\frac{i\,e^{2}k_{B}T}{\pi\hbar^{2}\!\left(\omega+i\tau^{-1}\right)}
\left[
\frac{\mu_c}{k_{B}T}
+2\ln\!\left(\exp\!\left(-\frac{\mu_c}{k_{B}T}\right)+1\right)
\right]
\tag{1.2}\\[4pt]
\sigma_{\mathrm{inter}}(\omega,\mu_c,\tau)
&=\frac{i\,e^{2}}{4\pi\hbar}
\ln\!\left[
\frac{2|\mu_c|-\hbar\left(\omega+i\tau^{-1}\right)}
     {2|\mu_c|+\hbar\left(\omega+i\tau^{-1}\right)}
\right]
\tag{1.3}
\end{align}

where $\sigma_{\mathrm{intra}}$ is the intraband term, $\sigma_{\mathrm{inter}}$ is the interband term, $\omega$ is the angular frequency, $\mu_c$ is the chemical potential, $T$ is the temperature, and $\tau$ is the graphene carrier scattering time. For 2D material graphene, $\mu_c$ can be regarded as the Fermi level, which is modulated by the applied voltage as:
\begin{equation}
\mu_c=\operatorname{sgn}(V)\,\hbar v_F
\sqrt{\frac{\pi V \varepsilon_0 \varepsilon_{\mathrm{Al_2O_3}}}{d_{\mathrm{Al_2O_3}}}}
\tag{1.4}
\end{equation}

Here $v_F$ is the Fermi velocity, $\varepsilon_0$ is the vacuum permittivity, $\varepsilon_{\mathrm{Al_2O_3}}$ is the dielectric constant of $\mathrm{Al_2O_3}$, and $d_{\mathrm{Al_2O_3}}$ is the thickness of the $\mathrm{Al_2O_3}$ layer. Combining Eqs.~(1.1)--(1.4), we obtain:
\begin{align}
\sigma_{\mathrm{intra}}(\omega,V,\tau)
&=\frac{i\,e^{2}k_{B}T}{\pi\hbar^{2}\!\left(\omega+i\tau^{-1}\right)}
\Bigg[
\frac{\operatorname{sgn}(V)\hbar v_F}{k_{B}T}
\sqrt{\frac{\pi V \varepsilon_0 \varepsilon_{\mathrm{Al_2O_3}}}{d_{\mathrm{Al_2O_3}}}}
\nonumber\\
&\hspace{3.5em}
+2\ln\!\left(
\exp\!\left(
-\frac{\operatorname{sgn}(V)\hbar v_F}{k_{B}T}
\sqrt{\frac{\pi V \varepsilon_0 \varepsilon_{\mathrm{Al_2O_3}}}{d_{\mathrm{Al_2O_3}}}}
\right)+1
\right)
\Bigg]
\tag{1.5}\\[6pt]
\sigma_{\mathrm{inter}}(\omega,V,\tau)
&=\frac{i\,e^{2}}{4\pi\hbar}
\ln\!\left[
\frac{
2\left|\operatorname{sgn}(V)\hbar v_F\sqrt{\frac{\pi V \varepsilon_0 \varepsilon_{\mathrm{Al_2O_3}}}{d_{\mathrm{Al_2O_3}}}}\right|
-\hbar\left(\omega+i\tau^{-1}\right)
}{
2\left|\operatorname{sgn}(V)\hbar v_F\sqrt{\frac{\pi V \varepsilon_0 \varepsilon_{\mathrm{Al_2O_3}}}{d_{\mathrm{Al_2O_3}}}}\right|
+\hbar\left(\omega+i\tau^{-1}\right)
}
\right]
\tag{1.6}
\end{align}

As indicated, the $\sigma_{\mathrm{intra}}$ and $\sigma_{\mathrm{inter}}$ are voltage determined, which is the mechanism for the E/O modulation.

\subsection{Graphene-silicon slot waveguide eigenmode and electro-absorption modulation }

The graphene--silicon waveguide absorption coefficient and refractive index can be calculated from the eigenmode of the graphene--silicon waveguide by employing the surface conductivity model in the simulation software Lumerical.

The finite-element method is used to calculate the optical eigenmode of the graphene-loaded device. The complex effective index $n_{\mathrm{eff}}$ is:
\begin{equation}
n_{\mathrm{eff}} = n_{\mathrm{eff,real}} + i\,n_{\mathrm{eff,imag}}
\tag{2.1}
\end{equation}

Here, $n_{\mathrm{eff,real}}$ is the refractive index of the silicon-slot waveguide E/O modulator. The corresponding propagation loss can be obtained from the imaginary part of the effective index:
\begin{equation}
\alpha~(\mathrm{dB}/\mu\mathrm{m})
=
\frac{2\pi}{\lambda_{0}}\,n_{\mathrm{eff,imag}}\times 4.343\times 10^{-6}
\tag{2.2}
\end{equation}

where $4.343\times 10^{-6}$ is the conversion constant from $\mathrm{m}^{-1}$ to $\mathrm{dB}/\mu\mathrm{m}$. Both the electro-refractive modulation and the electro-absorption modulation are voltage dependent.

Using the model and methods mentioned above, we calculate the graphene-loaded slot waveguide absorption coefficient and refractive index change versus applied voltage at the wavelength of 2\,\textmu m waveband. In the simulation, the graphene-loaded slot waveguide was assumed to have the graphene carrier scattering time of 10 fs and the dielectric layer thickness of 50 nm, and the graphene is undoped.

\begin{figure*}[h]
  \centering
  \includegraphics[width=0.40\linewidth]{s0.png}
  \caption{Effective absorption coefficient change versus the applied voltage for double layer graphene slot waveguide.}
  \label{fig:1}
\end{figure*}

Figure~\ref{fig:1} shows the graphene-silicon slot waveguide absorption coefficient and refractive index change versus the applied voltage. The graphene was assumed to have no pre-doping and show intrinsic properties. In the low voltage region, the inter-band transition plays the dominant role and graphene shows high absorption. In the high voltage region, the inter-band transition gradually vanishes, and the intra-band transition plays the dominant role. In this region, graphene becomes transparent for the incident photon.

\newpage
\section{Modeling Graphene-silicon slot waveguide}

\subsection{Graphene-silicon slot waveguide dielectric layer thickness optimization}

The dielectric layer thickness between the top and bottom graphene layers affects both the modulation bandwidth and the modulation efficiency of the modulator. A thicker dielectric layer can reduce the overlap capacitance of the double-layer graphene and thus increase the bandwidth; however, it also reduces the mode overlap between graphene and the optical field, leading to lower modulation efficiency. Therefore, there is a trade-off between modulation bandwidth and efficiency.

\begin{figure*}[h]
  \centering
  \includegraphics[width=0.4\linewidth]{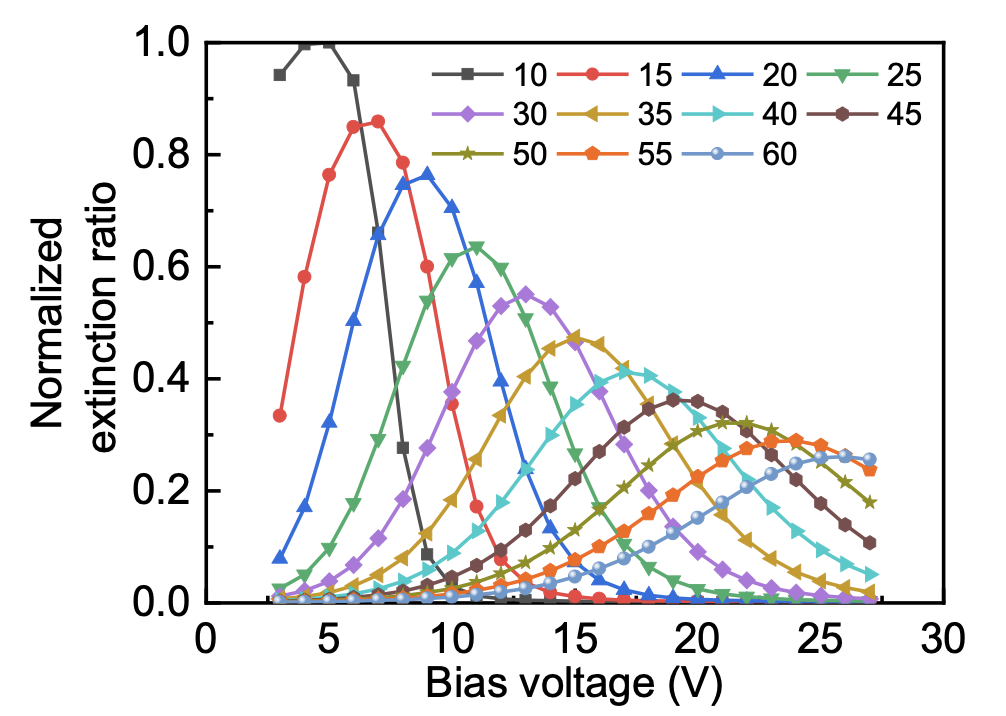}
  \caption{Normalized extinction ratio of the silicon-slot waveguide E/O modulator.}
  \label{fig:2}
\end{figure*}

We first calculate the absorption coefficient of the graphene--silicon slot waveguide, and then obtain the normalized extinction ratio (proportional to the modulation efficiency) for different dielectric layer thicknesses (assuming a driving voltage of \SI{6}{\volt}), as shown in Fig.~\ref{fig:2}.

\begin{figure*}[h]
  \centering
  \includegraphics[width=0.40\linewidth]{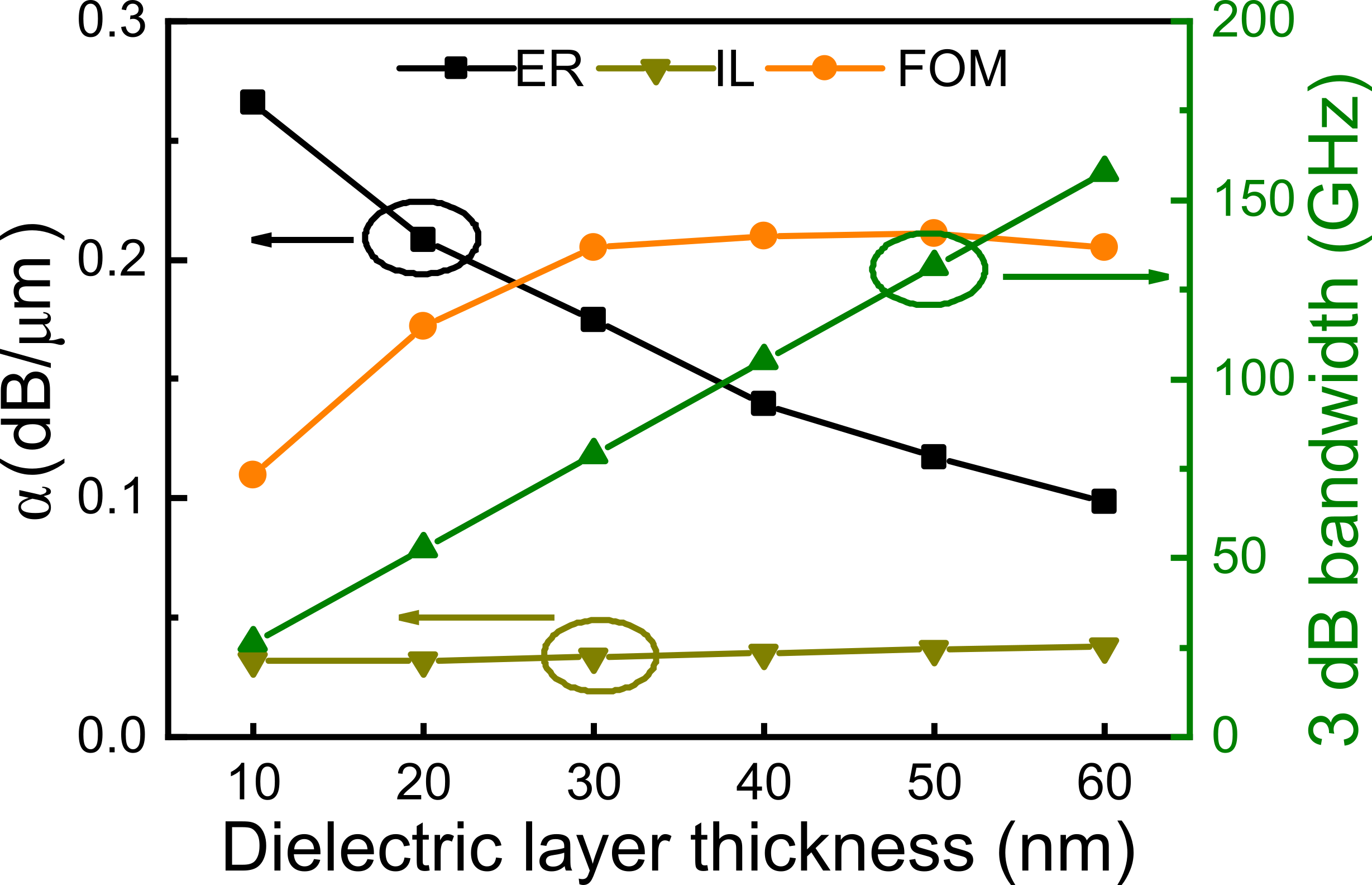}
  \caption{Modulation bandwidth, modulation efficiency, insertion loss, and the calculated modulation bandwidth--efficiency product as a function of dielectric layer thickness for a double-layer graphene-integrated microring modulator.}
  \label{fig:3}
\end{figure*}

Figure~\ref{fig:3} shows the normalized bandwidth and normalized modulation efficiency as a function of dielectric layer thickness. The bandwidth is calculated based on the device RC model. The figure of merit (FOM) is defined as the modulation bandwidth--efficiency product. The highest FOM is obtained for a modulator with a dielectric layer thickness of \SI{50}{\nano\meter}.

\subsection{Graphene-silicon slot waveguide double-layer graphene overlap width optimization}

\begin{figure*}[h]
  \centering
  \includegraphics[width=0.4\linewidth]{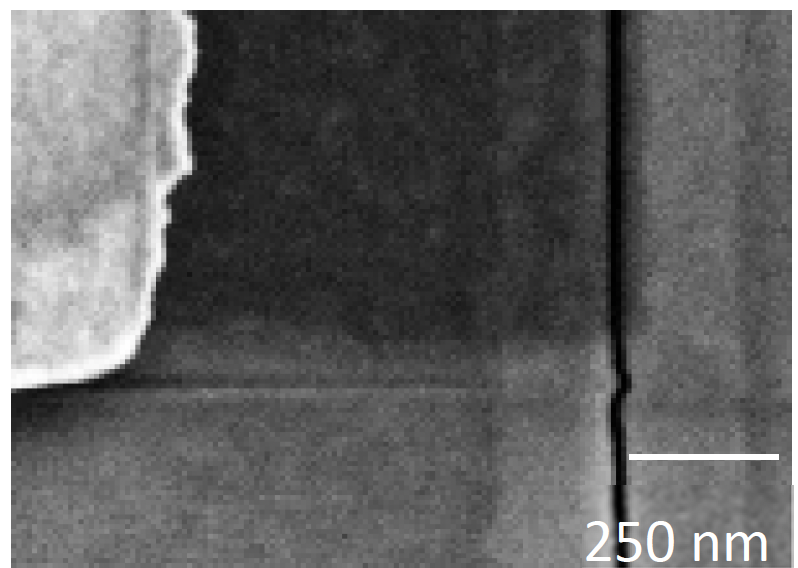}
  \caption{SEM image of the transferred first graphene layer, showing a high e-beam alignment accuracy of \SIrange{20}{30}{\nano\meter}.}
  \label{fig:4}
\end{figure*}

A smaller double-layer overlap leads to a larger bandwidth and a higher modulation efficiency--bandwidth product. Ideally, the overlap width should be equal to the slot width. In our experiment, a double-layer graphene overlap width of \SI{110}{\nano\meter} was used for a \SI{50}{\nano\meter}-wide slot waveguide, which is limited by the e-beam alignment accuracy. Fig.~\ref{fig:4} shows the SEM image after the transfer and patterning of the first graphene layer. The graphene is in contact with the waveguide on the opposite side of the slot, with an overlap width of approximately \SI{20}{\nano\meter} to \SI{30}{\nano\meter}.

Other parameters optimization strategy, including the slot waveguide dimension optimization, the electrode to waveguide distance optimization, and the electrical optimization of the modulator (electrical capacitance, resistance) are mentioned in the main manuscript and the response to the reviewers.

\newpage
\textcolor{black}{
\section{Graphene-silicon slot-waveguide electro absorption modulator fabrication}}

\textcolor{black}{
\subsection{Main fabrication process flow}}

\textcolor{black}{The fabrication starts with commercial SOI wafers with a 220-nm-thick silicon layer on top of a 2~\textmu m-thick SiO$_2$ buried layer. The wafer has high resistivity to reduce the capacitance. E-beam lithography and inductively coupled plasma etching are used to fabricate the slot waveguides and other passive components such as grating couplers, mode converters, and stripe waveguides. CSAR resist was adopted since it is a positive e-beam resist with high sensitivity, high resolution, and high plasma-etching resistance, making it suitable for defining large-scale devices with fine structures such as photonic-crystal grating holes and slot waveguides. The height difference between the fully etched silicon and the bottom buffer layer is 220~nm. To ensure a high graphene-transfer yield, PECVD SiO$_2$ was deposited and planarized to the top surface of the waveguide to provide a flat surface; otherwise, the graphene tends to break across the waveguide edge during drying. A spacer of about 5--7~nm-thick Al$_2$O$_3$ was then uniformly deposited on the waveguide surface by atomic layer deposition to avoid electrical connectivity between the graphene and the silicon waveguide and act as the SiO$_2$ etch stop layer.}

\begin{figure*}[h]
  \centering
  \includegraphics[width=0.80\linewidth]{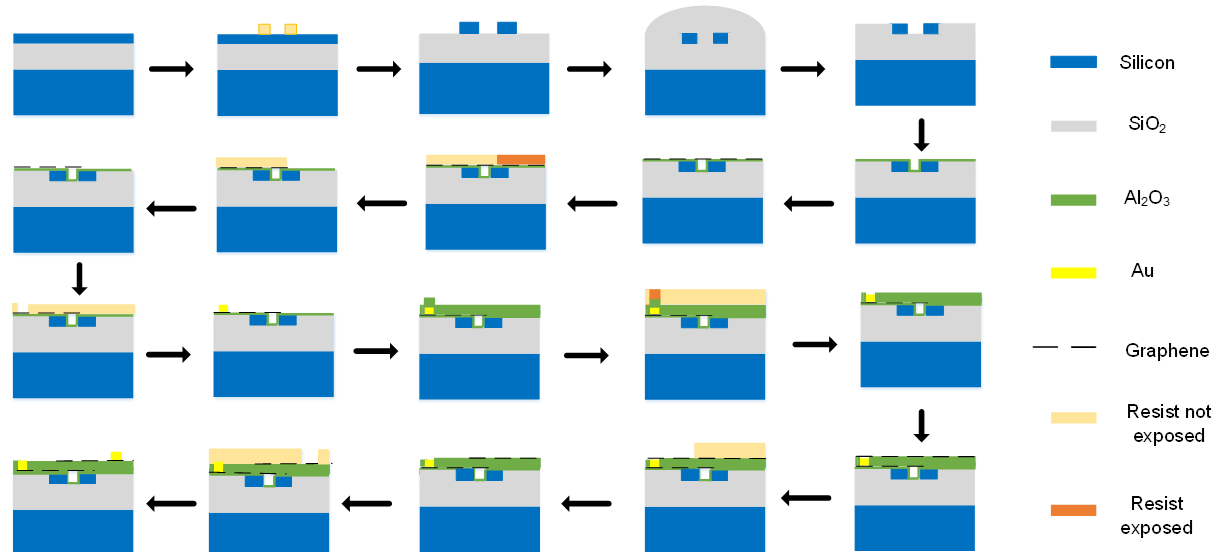}
  \caption{Graphene-silicon slot waveguide modulator fabrication process flow.}
  \label{fig:5}
\end{figure*}

\begin{figure*}[h]
  \centering
  \includegraphics[width=0.60\linewidth]{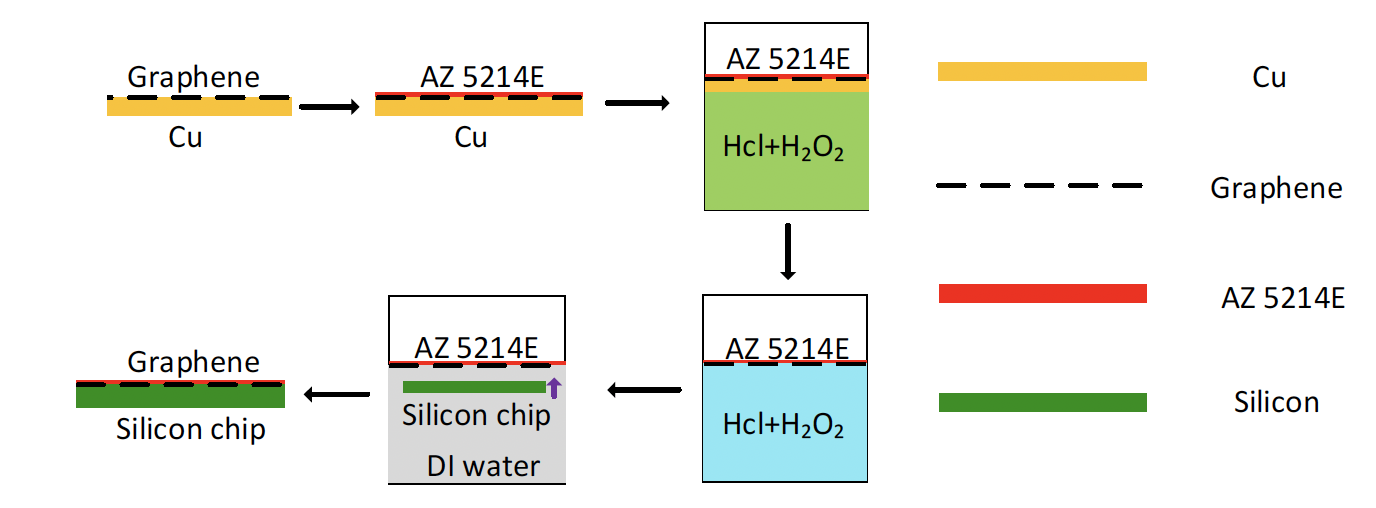}
  \caption{Large scale graphene wet transfer process.}
  \label{fig:6}
\end{figure*}

\textcolor{black}{CVD graphene on copper foil was spin-coated with 1~\textmu m AZ5214E resist, baked at \SI{90}{\degreeCelsius} for 2~minutes until dry, then soaked in a homemade copper etching solution (hydrochloric acid:DI water = 1:7, with a few drops of hydrogen peroxide) for 24~h and rinsed thoroughly in deionized (DI) water. To remove metal residuals, a diluted RCA clean step-two solution was employed before the wet-transfer process. The transferred graphene was left to dry for one week. The graphene pattern was then defined using e-beam lithography and O$_2$ plasma etching. Next, the metal pads were defined using UV lithography and deposited by e-beam evaporation following a lift-off process with 10~nm Ni and 100~nm Au. A rapid thermal annealing (RTA) process was then used to significantly improve graphene--pad conductivity and reduce contact resistance. During the RTA process, the sample was ramped to \SI{450}{\degreeCelsius} in \SI{30}{\second} and held at \SI{450}{\degreeCelsius} for approximately \SI{1}{\minute} under a flowing gas mixture of 10\% hydrogen in nitrogen, with the cycle repeated five times.}

\textcolor{black}{Direct deposition of high-dielectric-constant materials on pristine graphene by ALD is challenging due to the hydrophobic nature of the graphene basal plane. A 1~nm Al seed layer was thermally evaporated and quickly oxidized into Al$_2$O$_3$ upon exposure to air, after which 35~nm Al$_2$O$_3$ was deposited at \SI{200}{\degreeCelsius} by ALD. A similar process was performed for the top graphene layer and top metal pad as for the bottom layer. The top Al$_2$O$_3$ layer above the grating and metal pad was opened via H$_3$PO$_4$ wet etching. The complete process flow is shown in Figure~\ref{fig:5} and Figure~\ref{fig:6}.}

\textcolor{black}{
\subsection{Large aspect ratio slot waveguide fabrication}}

\textcolor{black}{The width of the slot waveguide is only \SI{50}{\nano\meter} while the etching depth of this slot is \SI{220}{\nano\meter}, it's pretty difficult to obtain a vertical etching side wall. Before etching the real device, we did an etching test to obtain the accurate etching speed as well as the etching profile.}

\begin{figure*}[h]
\centering
\includegraphics[width=0.6\textwidth]{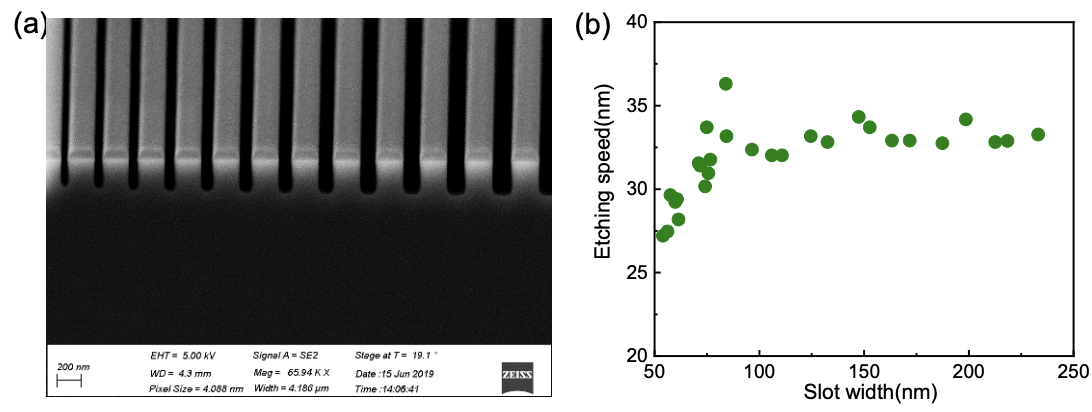}%
\caption{Etching profile and etching speed of the ASE with different slot width.}
\label{fig:7}
\end{figure*}

\begin{figure*}[h]
\centering
\includegraphics[width=0.6\textwidth]{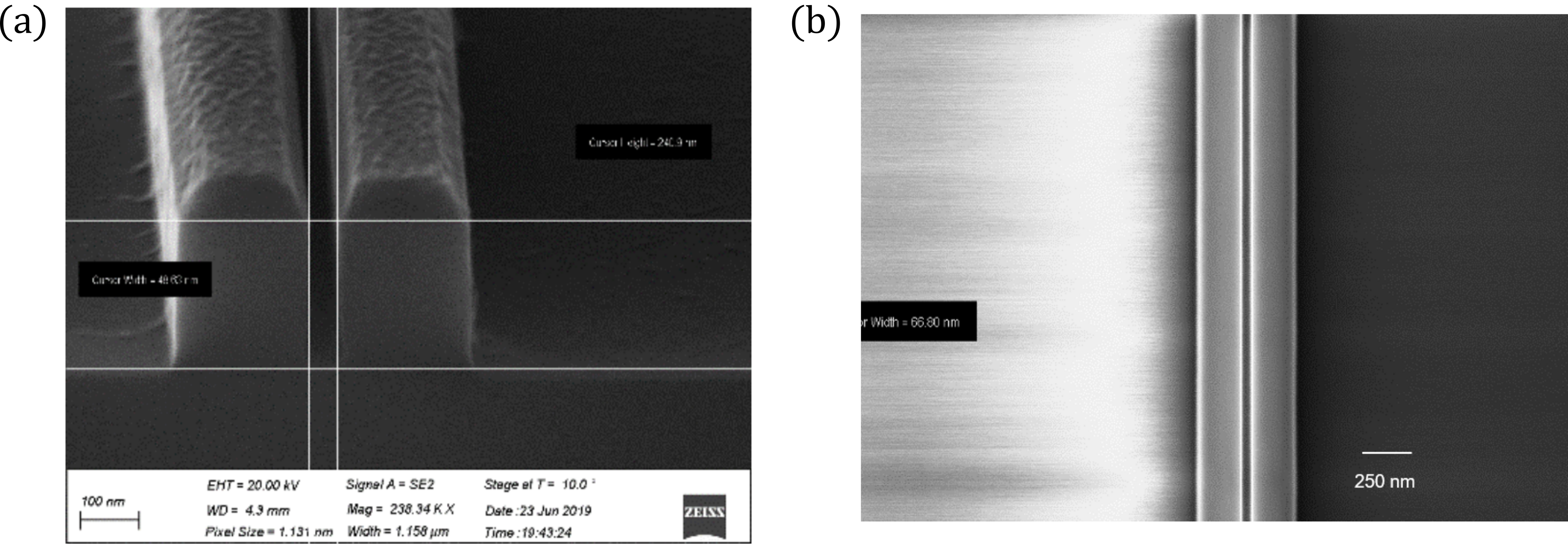}
\caption{Fabricated silicon slot waveguide with resist residual.}
\label{fig:8}
\end{figure*}

\textcolor{black}{Fig.~\ref{fig:7}(a) shows the etching profile and etching speed for the slot waveguide with different slot widths in the ASE machine. The etching speed increased when the slot width expanded from \SI{40}{\nano\meter} to \SI{100}{\nano\meter} and then the etching speed keeps at a constant speed of \SI{33}{\nano\meter} per etching cycle. Fig.~\ref{fig:8}(a) shows the fabricated silicon slot waveguide, the fabricated slot waveguide has a smooth sidewall and high aspect ratio.}

\textcolor{black}{
\subsection{Photonic crystal based apodized grating coupler}}

\begin{figure*}[h]
\centering
\includegraphics[width=0.5\textwidth]{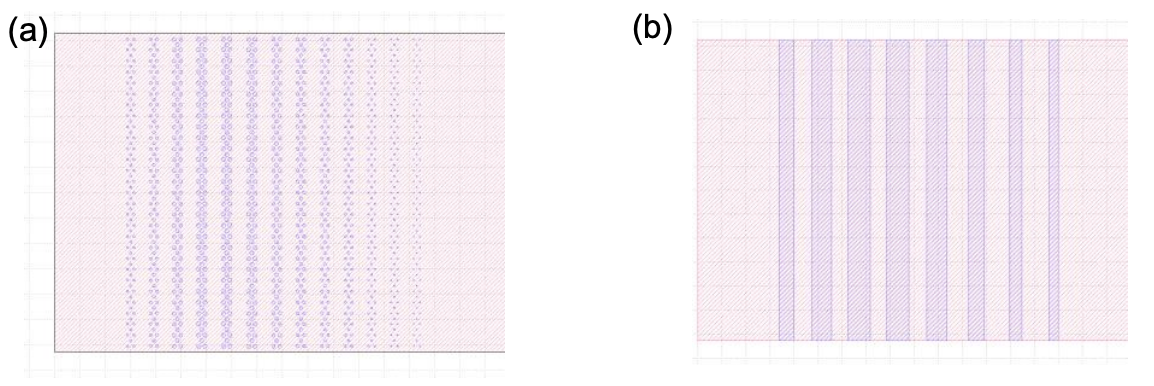}
\caption{Typical grating couplers with (a) grating coupler and (b) rectangular groves as element factor.}
\label{fig:9}
\end{figure*}

\textcolor{black}{The grating coupler could convert the light between the vertical fiber direction and the chip plane by periodical grating structures. Fig.~\ref{fig:9} shows two types of typical grating couplers consist of wide waveguide with periodical and uniform etched grooves. The grating equation which describes the coupled angle of the fiber ($\theta$) and the pitch of the grating ($\Lambda$), namely the Bloch equation is:}

\textcolor{black}{
\begin{equation}
(n_1 \sin\theta - n_{eff})\Lambda = m\lambda, \quad m = 0, 1, 2, \ldots
\label{eq:3.3.1}
\tag{4.1}
\end{equation}}

\textcolor{black}{Where $n_{eff}$ is the effective index of the grating and $n_1$ is the refractive index of the free space. This equation could also be regarded as the phase-matching condition in the grating, where the refractive index of the incident light that is projected into the horizontal direction should equal the eigenmode refractive index in the periodical grating structures. This equation must be fulfilled in order to have a high coupling efficiency.}

\textcolor{black}{For grating coupler working at TE mode the effective index is much higher than the air. So, in order to avoid the reflection loss caused by the large refractive index contrast, the grating equation requires that the etched depth of the grating trench has to be shallower than the waveguide in order to reduce the effective refractive index contrast. A standard etching depth of the grating in the foundry is \SI{70}{\nano\meter}, while the waveguide thickness is \SI{220}{\nano\meter}. The fabrication of the grating coupler thus involves two lithography and etching steps where critical alignment is needed. In order to simplify the fabrication process, we used a grating coupler based on the photonic crystal structure, which only needs one deep etching step. The photonic crystal structure also enables the refractive index engineering of the grating and tailor of the mode profile to fit the Gaussian shape of the standard single-mode fiber by periodical change the hole radius. This type of grating coupler is called apodized grating coupler, which has a higher coupling efficiency and less reflection than the uniform grating coupler.}

\textcolor{black}{The working principle and the design details of the apodized grating coupler based on photonic crystal are explained as follow:}

\textcolor{black}{The guided power along the grating is gradually decreasing due to the radiation of the grating into free space. The first step is to map the radiation efficiency with the effective refractive index and pitch of the grating cells.}

\textcolor{black}{The commonly used grating cells include many varieties, such as photonic crystal holes, rectangular holes, or round holes. These kinds of grating cells are subwavelength structures, of which the feature size is significantly smaller than the wavelength. They can be treated as an effective grating stripe with an engineer-able refractive index and leakage factor.}

\begin{figure*}[h]
\centering
\includegraphics[width=0.65\textwidth]{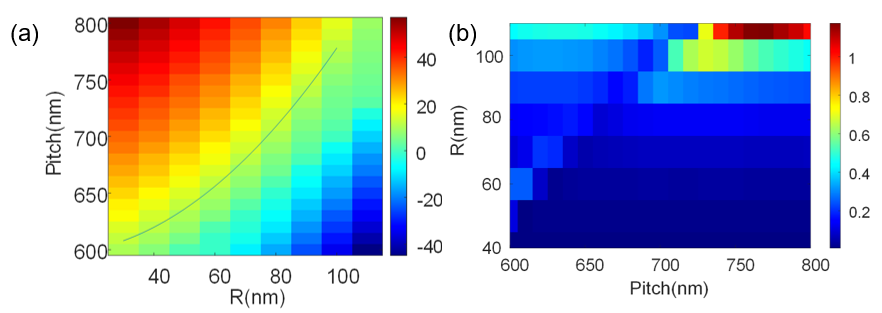}

\includegraphics[width=0.55\textwidth]{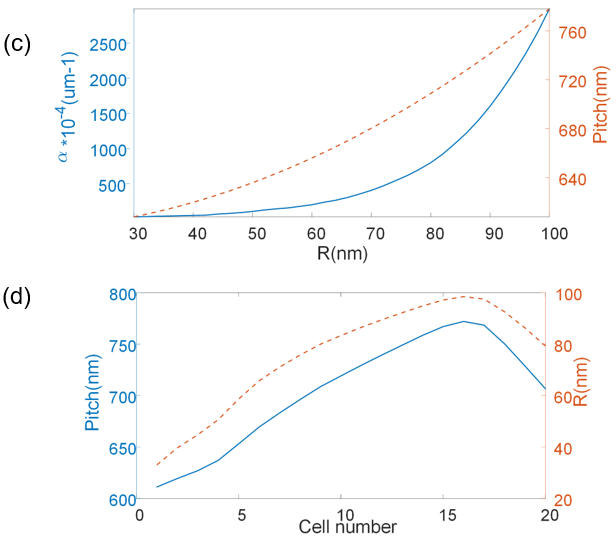}
\caption{(a) The emission angle of grating versus the photonic crystal radius and pitch. (b). The leakage factor of the grating versus the photonic crystal radius and pitch. (c). The parameters of the candidate grating cells with the corresponding leakage factor at angle of \SI{8}{\degree}. (d). The simulated parameters for the grating cells for the designed apodized grating coupler.}
\label{fig:10}
\end{figure*}

\textcolor{black}{Our design uses the photonics crystal in which small round holes are put in a triangle grid as shown in Fig.~\ref{fig:10}~(a). The diffraction angle and the leakage factor of the grating cell are first simulated in a uniform grating by 3D FDTD. The diffraction angle and the leakage factor versus the radius of the holes and the grating pitch are shown in Fig.~\ref{fig:10}~(b). Under the same pitch, the grating cell with a smaller radius has a large filling factor of silicon and results in a large effective refractive index but a smaller refractive index contrast within the cell. Therefore, a smaller radius corresponds to a large diffraction angle and a weak leakage factor. Combined with Eq.~\ref{eq:3.3.1}, a large grating pitch has a large diffractive angle and weak leakage factor.}

\textcolor{black}{Based on Fig.~\ref{fig:10}~(a), we can fit a line of equal diffraction angle on the map.}

\textcolor{black}{The corresponding radiuses and the grating pitches are the parameters for the candidate grating cells for desired diffraction angle. A diffraction angle of \SI{8}{\degree} was selected as this is the commonly used angle in commercial products. If one can customize the diffraction angle, the selection of angle depends on the thickness of the bottom oxide layer, in which the light transmitted downwards and the light reflected from the substrates interfere with each other. The ideal angle is what corresponds to constructive interference. The leakage factor of the grating cells on the line of equal diffraction angle can be fitted from the map of Fig.~\ref{fig:10}~(b). The parameters of the candidate grating cells that correspond to the desired radiation power at the desired angle are summarized in Fig.~\ref{fig:10}(c). The last step is to fit from the candidate grating cells to form a leakage factor distribution as in Fig.~\ref{fig:10}(c) which has a Gaussian-like radiation field, namely the apodized grating. The radiuses and grating pitches along the target grating are shown in Fig.~\ref{fig:10}~(d). The mathematic expression behind the 3D FDTD simulation is as follows:}

\textcolor{black}{The guided power $A^2_{(z)}$ along the grating is gradually decreasing due to the radiation of the grating into the free space of which the power is $B^2_{(z)}$. The leakage factor of the grating along the propagation direction is determined by the radiation field $B^2_{(z)}$: }

\textcolor{black}{
\begin{equation}
\alpha(z) = \frac{B^2(z)}{2\left[A_0^2 - \int_0^z B^2_{(t)} dt\right]}
\label{eq:3.3.2}
\tag{4.2}
\end{equation}}

\textcolor{black}{
The radiation field $B^2_{(z)}$ should match with the mode profile of the single-mode fiber, which is approximated as a Gaussian function:} 

\textcolor{black}{
\begin{equation}
B_{(z)} = B_0 \exp\left(-\frac{(z - z_m)^2}{2\sigma_g^2}\right)
\label{eq:3.3.3}
\tag{4.3}
\end{equation}}

\textcolor{black}{Where $B_0$ is the maximal value of the field distribution and is determined by the difference of the $A^2_{(z)}$ at the beginning and the end. Substituting the Gaussian radiation field into Eq.~\ref{eq:3.3.2}, we can have the leakage factor distribution $\alpha(z)$ as depicted in Fig.~\ref{fig:10}(c).}

\textcolor{black}{Simulations steps for the grating coupler works at \SI{2}{\micro\meter} wavelength are the same as the process for the \SI{1.5}{\micro\meter} wavelength, the only differences are the photonic crystal constant, photonic crystal pitch, photonic crystal radius, and the mode profile distribution in the standard single-mode fibers. Compared with \SI{1.5}{\micro\meter} wavelength grating, the above mentioned parameters for the \SI{2}{\micro\meter} wavelength grating should be multiplied by 1.3.}

\textcolor{black}{
\subsection{Rapid thermal annealing to reduce graphene-metal contact resistance}}

\textcolor{black}{Rapid thermal annealing (RTA) is a well-established technique for reducing graphene–metal contact resistance. To quantify its effect, we measured the graphene–metal contact resistance and graphene sheet resistance both before and after the RTA process.}

\begin{figure*}[h]
\centering
\includegraphics[width=0.55\textwidth]{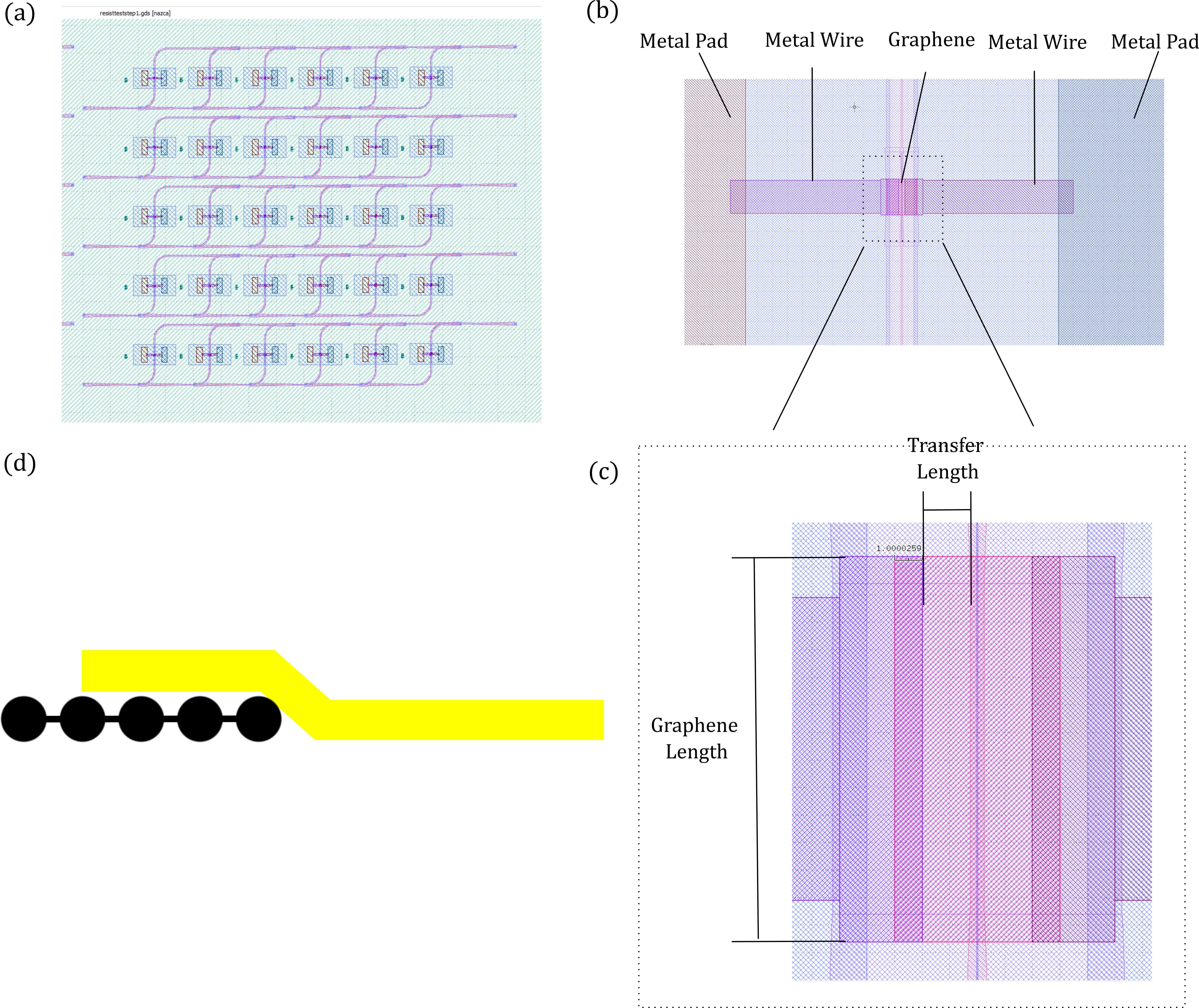}
\caption{\textcolor{black}{(a,b,c): Device pattern to measure the graphene-metal contact resistance and transfer resistance. (d): Schematic of the graphene-metal 1d-2d hybrid contact.}}
\label{fig:11}
\end{figure*}

\textcolor{black}{Fig.~\ref{fig:11} shows the GDS layout of the test structures fabricated for this characterization. To minimize contact resistance, we employed a Ni/Au metallization (10 nm Ni / 100 nm Au) with a hybrid contact scheme combining both one-dimensional edge and two-dimensional surface contact between the metal and graphene. The test structures were designed with varying graphene–metal contact lengths and transfer widths to enable reliable extraction of the contact and sheet resistance. The graphene–metal surface overlap was fixed at 1 $\mu$m—limited by the alignment accuracy of the UV maskless lithography system used to define the metal regions—in order to minimize parasitic capacitance in the contact area.}

\textcolor{black}{Prior to RTA, we extracted an average contact resistance of 3315~$\Omega{\cdot}\mu$m (the contact length contribution is neglected as it is negligible compared with the contact width) and a sheet resistance of 600~$\Omega/\square$. Following the RTA process, the contact resistance decreased substantially to 1050~$\Omega{\cdot}\mu$m---a reduction of approximately 68\%---while the sheet resistance increased only marginally to 670~$\Omega/\square$. These results confirm that RTA is highly effective at improving the graphene--metal interface with minimal degradation of the graphene sheet conductivity.}

\begin{figure*}[h]
\centering
\includegraphics[width=0.8\textwidth]{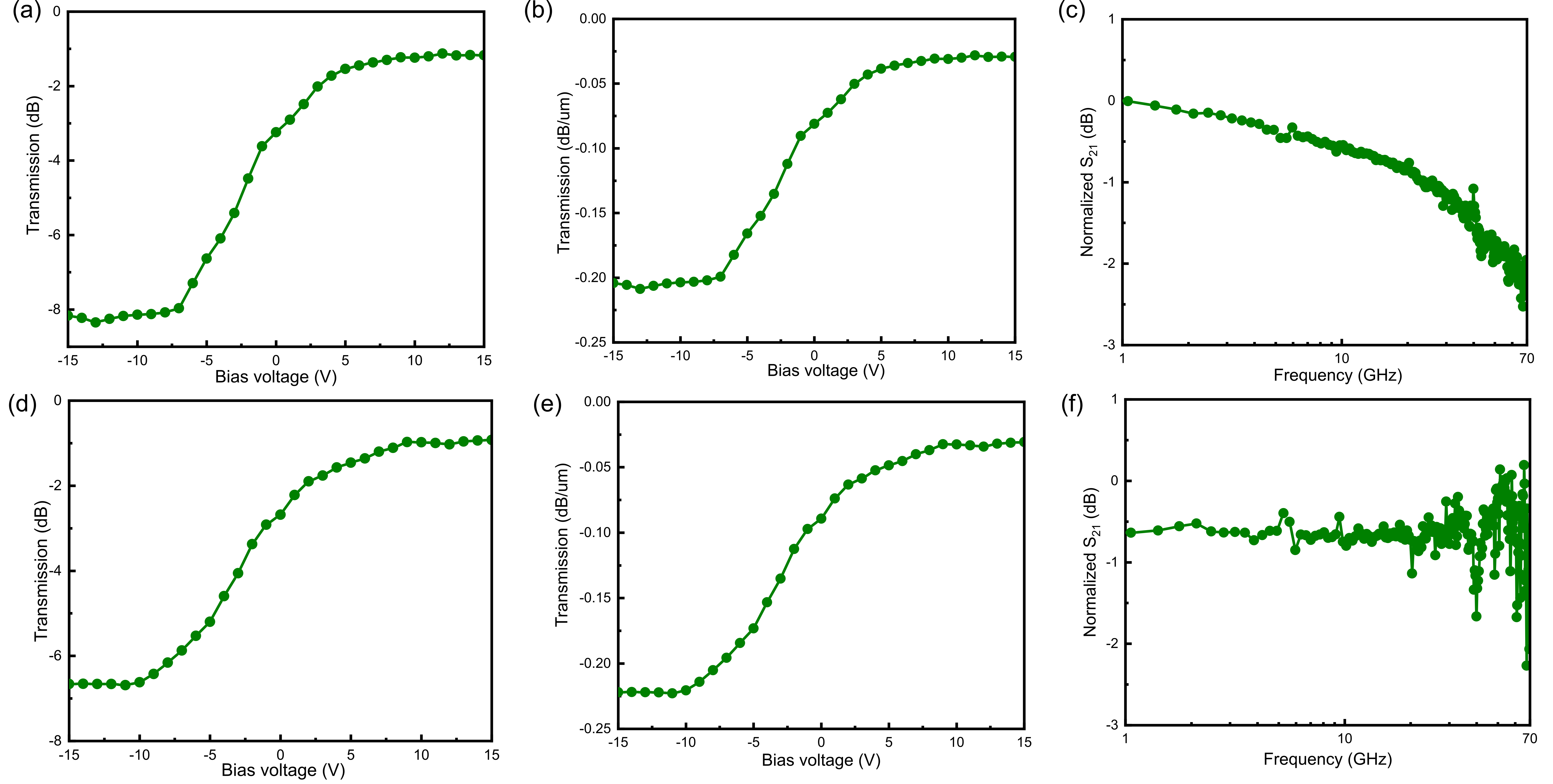}
\caption{\textcolor{black}{(a) Graphene-silicon slot waveguide modulator at 1.5~\textmu m waveband.}}
\label{fig:15}
\end{figure*}

We fit the capacitance and the resistance of the fabricated modulator using the measured $S_{11}$ data and the RC model of the device (Fig.~\ref{fig:15}). 

The relationship between the RC model and the measured $S_{11}$ data is as follows: 

\begin{equation}
Z_{in}(\omega) = Z_0 \frac{1 + S_{11}(\omega)}{1 - S_{11}(\omega)}
\label{eq:3}
\end{equation}

where $Z_0 = 50\,\Omega$. $S_{11}(\omega)$ is frequency dependent and was measured by VNA.

The impedance could be divided into three parts as below:

\begin{equation}
Z_{left}(\omega) = \frac{1}{j\omega C_{air}}
\end{equation}

\begin{equation}
Z_{medium}(\omega) = \frac{1}{j\omega C_g} + 2\cdot R_c + 2\cdot R_s
= \frac{1}{j\omega C_g} + 2\cdot R_g
\end{equation}

\begin{equation}
Z_{right}(\omega) = \frac{1}{j\omega C_{box}} + 2\cdot R_{sub}
\end{equation}

The total impedance $Z(\omega)$ is as below:

\begin{equation}
\frac{1}{Z_{in}(\omega)}
= \frac{1}{Z_{left}(\omega)} + \frac{1}{Z_{medium}(\omega)} + \frac{1}{Z_{right}(\omega)}
\end{equation}

\begin{equation}
= \frac{1}{\dfrac{1}{j\omega C_{air}}}
+ \frac{1}{\dfrac{1}{j\omega C_g} + 2\cdot R_g}
+ \frac{1}{\dfrac{1}{j\omega C_{box}} + 2\cdot R_{sub}}
\end{equation}

Taking the measured $S_{11}$ data into the equation, the fitted values are as follows:

\begin{table}[h]
\centering
\color{black}
\begin{tabular}{|c|c|c|c|c|c|}
\hline
$C_\text{air}$ & $C_\text{graphene}$ & $C_\text{box}$ & $R_\text{s}$ & $R_\text{c}$ & $R_\text{sub}$ \\
\hline
4~fF & 2~fF & 3.6~fF & 100~$\Omega$ & 200~$\Omega$ & 10000~$\Omega$ \\
\hline
\end{tabular}
\end{table}

\begin{figure*}[h]
\centering
\includegraphics[width=0.4\textwidth]{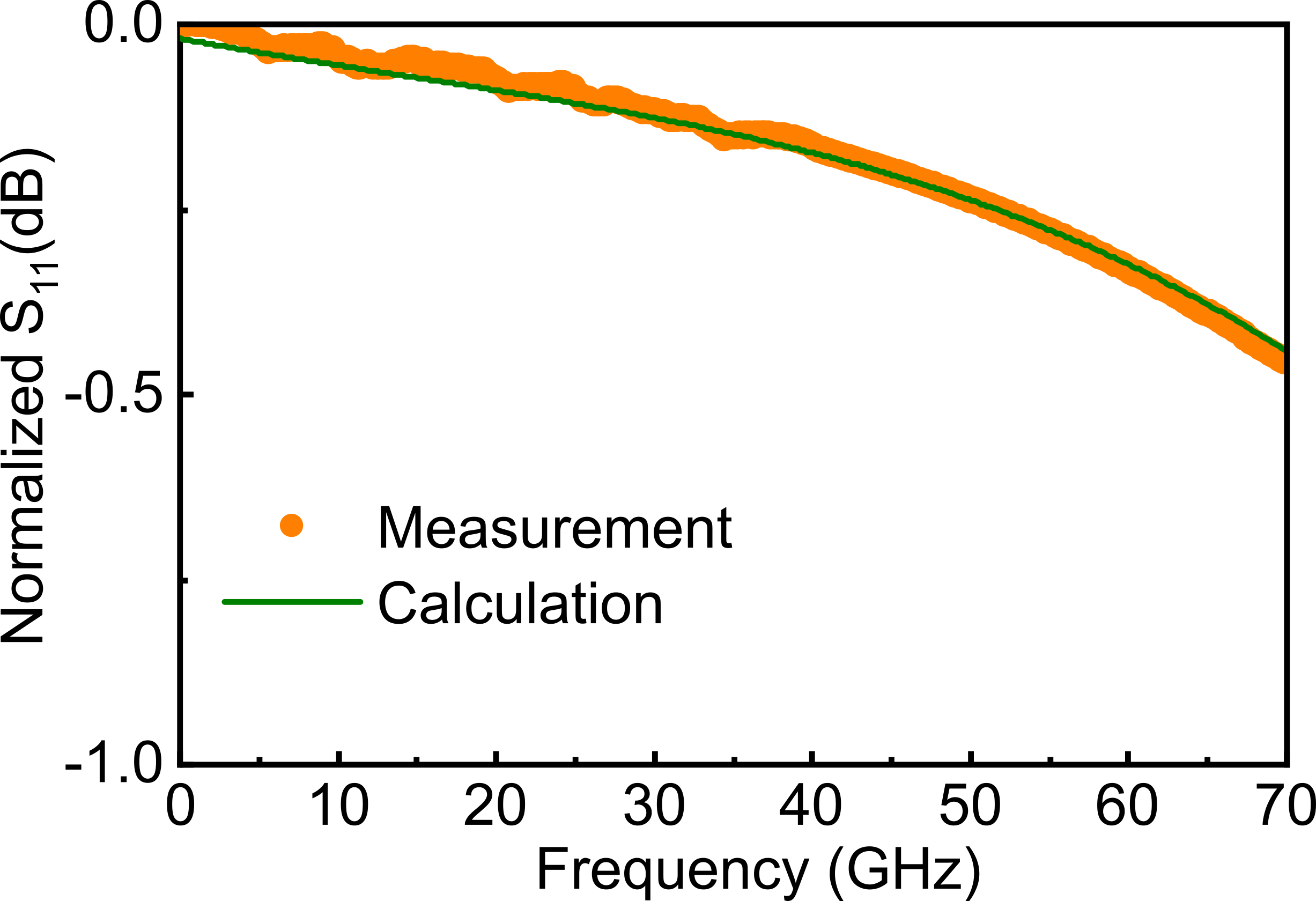}
\caption{\textcolor{black}{Measured and fitted device S$_{11}$ to 70 GHz.}}
\label{fig:12}
\end{figure*}

\begin{figure*}[h]
\centering
\includegraphics[width=0.4\textwidth]{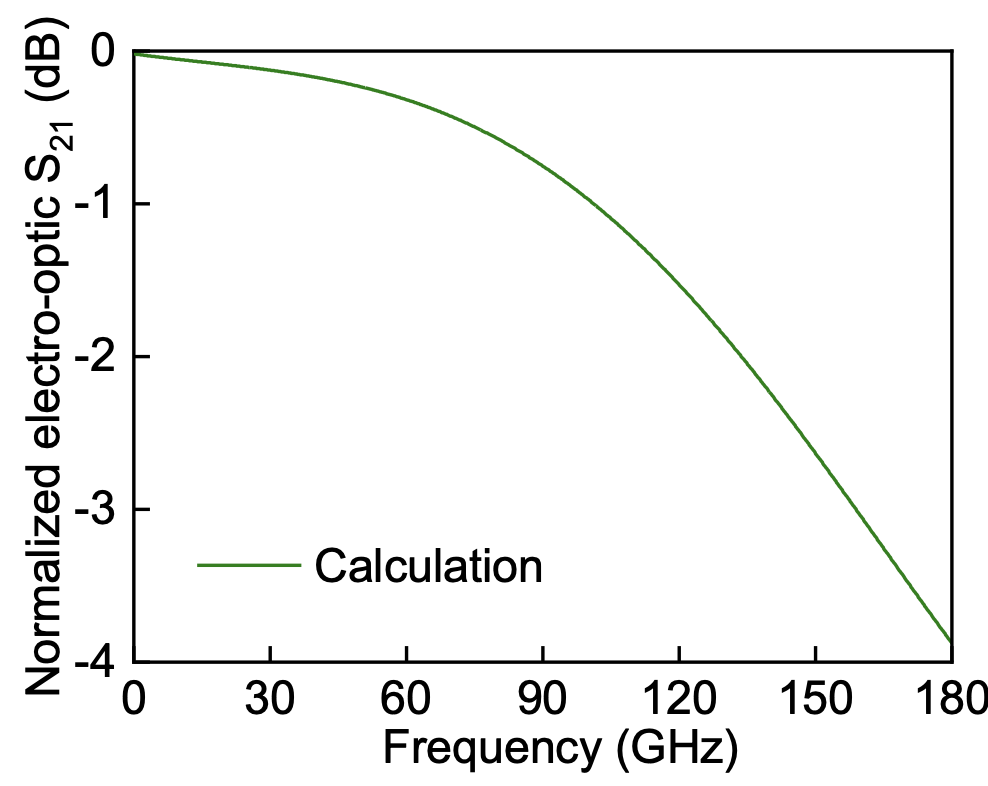}
\caption{\textcolor{black}{Calculated frequency response of the electro absorption modulator.}}
\label{fig:121}
\end{figure*}

\textcolor{black}{From the fitted values of $R_\text{s}$ and $R_\text{c}$, the graphene--metal contact resistance is calculated to be 1200~$\Omega{\cdot}\mu$m, and the graphene sheet resistance is estimated to be 860~$\Omega/\square$.}

\textcolor{black}{These values are moderately higher than those obtained from the dedicated test structures (1200~$\Omega{\cdot}\mu$m for contact resistance and 860~$\Omega/\square$ for sheet resistance). This discrepancy is primarily attributed to additional resist residues accumulated during the full device fabrication process, which involves a greater number of lithography and transfer steps than the simpler test structures used for standalone resistance characterization. Nevertheless, the fitted and directly measured values are in reasonable agreement, lending confidence to the extracted device parameters.}

\textcolor{Using the calculated capacitance and resistance, we calculate the frequency response of the modulator at \SI{2}{\micro\meter} wavelength band, as indicated in Fig.~\ref{fig:121}, the modulator has a large 3-dB bandwidth over 150 Ghz.}

\newpage
\textcolor{black}{
\section{Graphene-silicon slot-waveguide electro absorption modulator characterization}}

\textcolor{black}{\subsection{DC and high-speed characterization of the modulator}}

\begin{figure*}[h]
\centering
\includegraphics[width=0.6\textwidth]{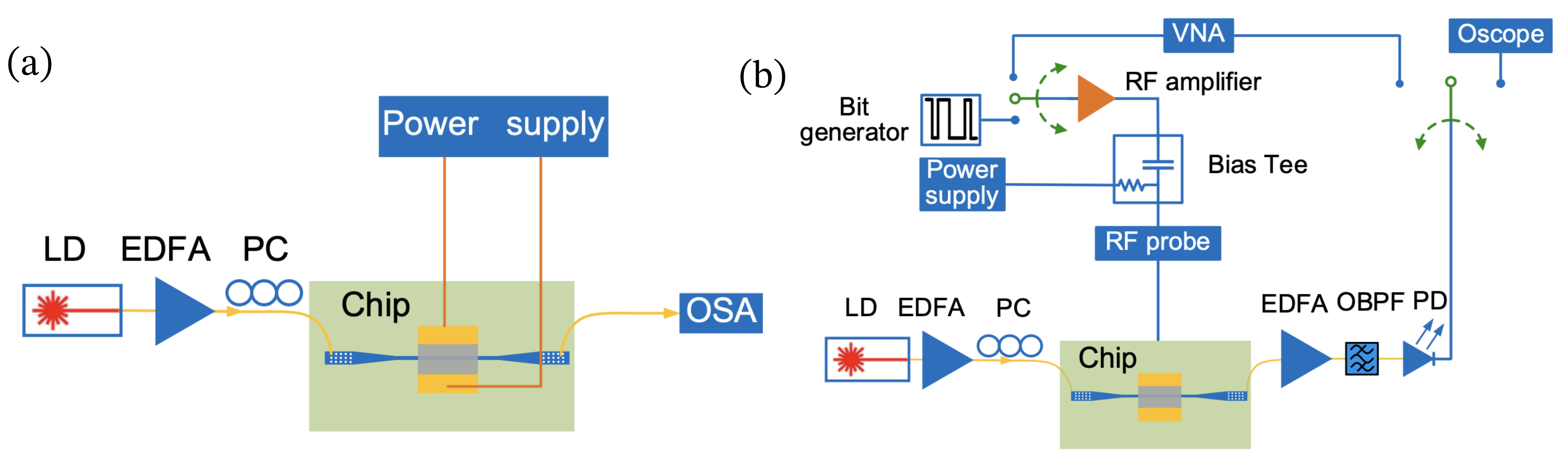}
\caption{\textcolor{black}{(a) DC and (b) high speed measurement setup of the modulator.}}
\label{fig:13}
\end{figure*}

\textcolor{black}{
Fig.~\ref{fig:13} illustrate the configurations for static and high-speed measurements of the graphene-silicon slot-waveguide electro-optic (E/O) modulator at the \SI{1.5}{\micro\meter} waveband. At a wavelength of \SI{2}{\micro\meter} waveband, we employ a Thulium-Doped Fiber Amplifier (TDFA) to amplify the incident optical power.} 

\textcolor{black}{
A small-signal driving voltage on the order of milliwatts from the VNA was first amplified to 2~V using the RF driver. A bias-tee was used to combine the amplified RF driving voltage and the d.c.\ bias voltage. Two needles of the high-speed RF probe station were carefully adjusted to horizontally contact the two pads of the modulator. The light coupled out from the chip was amplified by a preamplifier and then sent to a 70~GHz photodetector to generate a high-speed electrical signal, which was received by the VNA input port. The E/O modulator bandwidth was extracted by removing the RF power loss of the entire transmission system except the modulator, the measured bandwidth is over 70~GHz, which is the maximum bandwidth of our VNA.}

\textcolor{black}{
At \SI{2}{\micro\meter} waveband, similar methods was employed to measure the bandwidth before 20 GHz. Beyond 20~GHz, we adopted an optical sideband analysis technique to extend the characterization up to 40~GHz. In this approach, a single-frequency RF tone from the AWG was applied to the modulator at each frequency step, and the resulting modulated optical spectrum was recorded using a high-resolution optical spectrum analyser (OSA). The electro-optic response at each RF frequency was then extracted from the power ratio between the first-order modulation sideband and the optical carrier. The two measurement techniques yield consistent results in their overlapping frequency range near 20~GHz, validating the cross-calibration and confirming the continuity of the frequency response across the transition between methods. The combined dataset reveals a modulation bandwidth exceeding 40~GHz at the 2~\textmu m waveband, with no 3-dB roll-off observed within the measured range.}

\textcolor{black}{\subsection{Electro absorption modulator with different graphene length.}}

\begin{figure*}[h]
\centering
\includegraphics[width=0.8\textwidth]{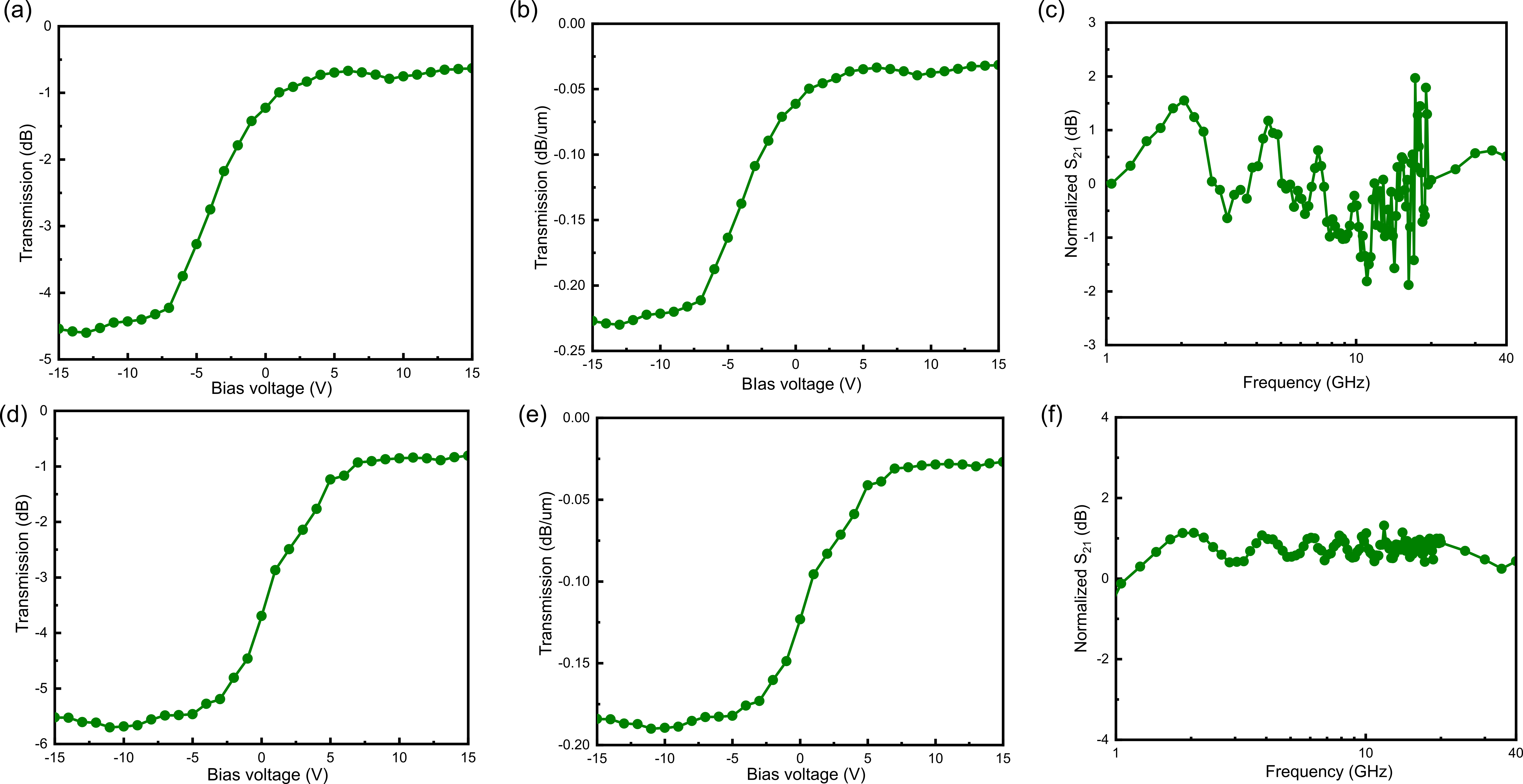}
\caption{\textcolor{black}{(a) Graphene-silicon slot waveguide modulator at 2~\textmu m waveband.}}
\label{fig:14}
\end{figure*}

The graphene-silicon slot waveguide electro absorption modulator with different graphene lengths were fabricated and characterized, as indicated in Fig.~\ref{fig:14}.

As indicated in Fig.~\ref{fig:15}, the graphene is p doped, which comes from the pre-fabrication doping from the graphene CVD process and the fabrication and wet transfer induced doping. Figure shows the Raman spectrum of the graphene before wet transfer, indicating the pre-doping condition of 0.3 eV.

\begin{figure*}[h]
\centering
\includegraphics[width=0.6\textwidth]{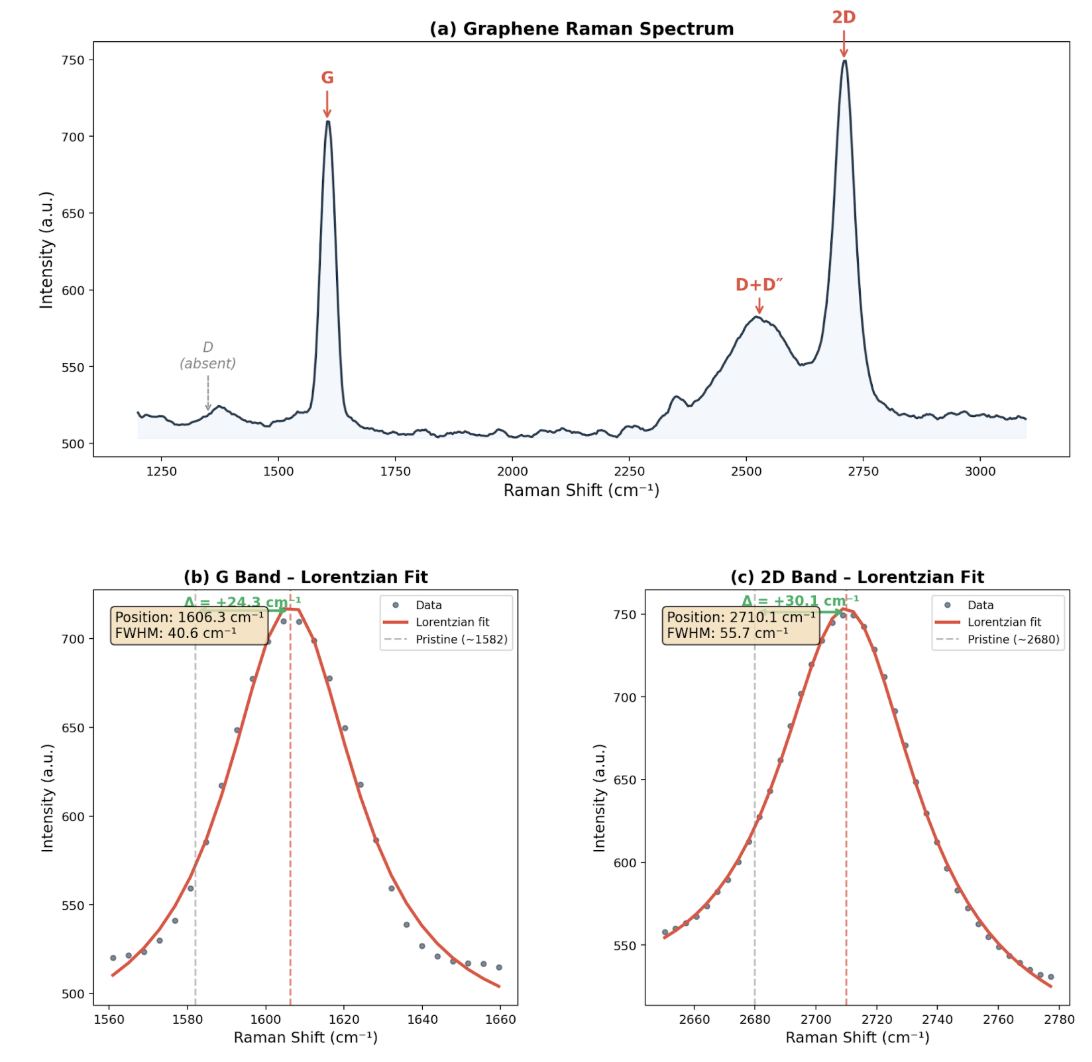}
\caption{\textcolor{black}{(a) Raman spectrum of the graphene.}}
\label{fig:16}
\end{figure*}

\newpage
\textcolor{black}{
\section{Advantages of the 2-\textmu m waveband communication}}

\textcolor{black}{
The 2-\textmu m band is currently a leading contender among all the new wavelength bands due to the combination of several emerging technologies. Firstly, the thulium-doped fibre amplifier (TDFA) offers more than 240-nm bandwidth in the 2-\textmu m band with high gain and low noise figure. Secondly, hollow-core fibre (HCF), e.g., using photonic bandgap or anti-resonant effects, enables data transmission in the 2-\textmu m band. HCF could potentially achieve a loss of the order of 0.1~dB/km, surpassing the best conventional silica fibres. Photonic bandgap fibres have been made with $>150$-nm bandwidth in the 2-\textmu m band and antiresonant fibres with 700-nm bandwidth covering both the conventional C band and the 2-\textmu m band. HCF also features low-latency transmission at $>99.7\%$ of the speed of light, and ultralow nonlinearity. Just as low-loss standard single-mode fibre (SSMF) and the erbium-doped fibre amplifier (EDFA) have boosted the optical fibre telecom industry, HCF and the TDFA could ultimately spawn novel 2-\textmu m-band optical communication systems, allowing signals to be relayed over long distances with low cost and low energy consumption. In addition, the 2-\textmu m band is attractive for silicon photonics due to the reduced two-photon absorption, enabling advanced on-chip functionalities.}

\newpage
\textcolor{black}{
\section{Performance comparison of on-chip integrated EO modulators operating at 2-\textmu m waveband}}

\begin{table*}[h]
\centering
\caption{Comparison of representative E/O modulators operating at the 2-\textmu m waveband.}
\renewcommand{\arraystretch}{2}
\setlength{\tabcolsep}{4pt}
\resizebox{\textwidth}{!}{%
\begin{tabular}{|c|c|c|c|c|c|c|}
\hline
\textbf{Ref.} & \textbf{Platform} & \textbf{Structure} & \textbf{Efficiency (dB/\textmu m)} & \textbf{Bandwidth (GHz)} & \textbf{Footprint (\textmu m)} & \textbf{Other wavebands capable} \\
\hline
\cite{cao2018highspeed2um} & SOI  & MZI            & 5.8/1500  & 10    & 1500  & C-band \\
\hline
\cite{cao2021high} & SOI  & Michelson      & 15/1500        & --    & 1500    & No \\
\hline
\cite{wang2021high} & SOI  & MZI            & 22/2000        & 18    & 2000    & No \\
\hline
\cite{shen2022bistability} & SOI  & MRM            & 15/62.8        & 18    & 62.8   & No \\
\hline
\cite{hagan2020tdfa} & SOI  & MRM            & 15/94.2        & 12.5    & 94.2   & No \\
\hline
\cite{wang2024efficient} & SOI  & Racetrack ring & 21/1000        & 26    & 1000    & No \\
\hline
\cite{pan2021lnoi2um} & LNOI & MZI            & 20/5000        & 22    & 5000    & No \\
\hline
\cite{li2026ultra} & LNOI & MZI            & 18/9000   & $>50$ & 9000  & O--U bands \\
\hline
This work & Graphene & Electro-absorption & 0.22 & $>40$ & 6 & C-band \\
\hline
\end{tabular}%
}
\end{table*}

Table~I summarizes the performance metrics of integrated electro-optic modulators operating in the 2-\textmu m spectral band. Our electro absorption modulator achieves a $>40$~GHz EO 3-dB bandwidth with a compact footprint of only 6-\textmu m and a high modulation efficiency of 0.22 dB/\textmu m. This advancement establishes a new benchmark for energy efficient high-speed data transmission in the 2-\textmu m band.

\newpage
\renewcommand{\bibsection}{\section*{REFERENCES}}
\bibliography{main}